\documentclass[tradiabstract]{aa} 

\usepackage{natbib}
\usepackage{graphicx}
\usepackage{placeins}
\usepackage{amsmath}
\usepackage{arydshln}
\usepackage{geometry}
\usepackage{xcolor}
\usepackage{placeins}

\begin{document} 

\title{Two-fluid simulations of Rayleigh-Taylor instability in a magnetized solar prominence thread II. Effects of collisionality.}
\author{B. Popescu Braileanu
          \inst{1,2,3}
                \thanks{ \email{\tiny beatriceannemone.popescubraileanu@kuleuven.be}}
          \and
          V. S. Lukin\inst{4}
          \and
          E. Khomenko\inst{1,2}
          \and
          \'A. de Vicente\inst{1,2}}
          
\titlerunning{The Rayleigh-Taylor instability II}
\authorrunning{Popescu Braileanu et al.}

\institute{Instituto de Astrof\'{\i}sica de Canarias, 38205 La Laguna, Tenerife, Spain
\and Departamento de Astrof\'{\i}sica, Universidad de La Laguna, 38205, La Laguna, Tenerife, Spain
\and Centre for mathematical Plasma-Astrophysics, Celestijnenlaan 200B, 3001 Leuven, KU Leuven, Belgium
\and National Science Foundation, Alexandria, VA, 22306, USA}

\date{Received 2021; Accepted XXXX}
 
\abstract{
Solar prominences are formed by partially ionized plasma with inter-particle collision frequencies generally warranting magnetohydrodynamic treatment.  In this work we explore the dynamical impacts and observable signatures of two-fluid effects in the parameter regimes when ion-neutral collisions do not fully couple the neutral and charged fluids. We perform 2.5D two-fluid (charges - neutrals) simulations of the Rayleigh-Taylor instability (RTI) at a smoothly changing interface between a solar prominence thread and the corona. The purpose of this study is to deepen our understanding of the RTI and the effects of the partial ionization on the development of RTI using nonlinear two-fluid numerical simulations. Our two-fluid model takes into account { neutral} viscosity, thermal conductivity, and collisional interaction between neutrals and charges: ionization--recombination, 
energy and momentum transfer, and frictional heating.  {In this paper (hereafter  paper II), the sensitivity of the RTI dynamics to collisional effects for different magnetic field configurations supporting the prominence thread is explored.  This is done by artificially varying, or eliminating, effects of both elastic and inelastic collisions by modifying the model equations.  We find that ionization and recombination reactions between ionized and neutral fluids  do not substantially impact the development of the primary RTI. However, such reactions can impact the development of secondary structures during the mixing of the cold prominence and hotter surrounding coronal material.  We find that collisionality within and between ionized and neutral particle populations plays an important role in both linear and nonlinear development of RTI;   ion-neutral collision frequency is the primary determining factor in development or damping of small-scale  structures.  We also observe that the degree and signatures of flow decoupling between ion and neutral fluids can depend  on the inter-particle collisionality and on the magnetic field configuration of the prominence thread.}
}

\keywords{Sun: chromosphere -- Sun: instabilities -- Sun: magnetic field -- Sun: numerical simulations}

\maketitle
%

\section{Introduction}

Solar prominences consist of plasma confined in large-scale magnetic structures. The prominence plasma is usually { about two orders of magnitude} cooler and denser than the surrounding corona, and it is believed to have chromospheric origin \citep{2010Labrosse}.  Therefore, in a prominence, plasma is only partially ionized, and the timescale associated with collisions between neutrals and ions can be comparable to  hydrodynamic timescales.
{ There have been a number of observational studies looking for indications of decoupling in velocity between ions and neutrals \citep{2007Gilbert, 2016Khomenko, 2017Anan}.}

The Rayleigh-Taylor instability (RTI) occurs when a heavier fluid is accelerated against a lighter fluid. In a two-dimensional geometry, a small perturbation  at the interface between the heavy and light fluids that varies in the direction perpendicular to the direction of the gravity may grow without bounds, depending on the scale of the perturbation.  It is well known  that within the ideal magnetohydrodynamics (MHD) model a discontinuous interface between heavy and light fluids is RT-unstable, and that the linear growth rate increases with increasing mode number \citep{Ch1961}.  However, when the viscosity is taken into account the growth rate achieves a maximum for a finite value of the mode number, and falls off asymptotically to zero for higher mode numbers \citep{Ch1961}. 









%

Analytical models that use more complex density profiles have also shown that the viscosity has a stabilizing effect on the linear growth rate of RTI \citep{1974Bhatia, Mikaelian1993}. \cite{Visco-nonlin} showed that viscous effects also inhibit the nonlinear development of the RTI by inhibiting the secondary Kelvin-Helmholtz instability (KHI).

In a laboratory plasma context, inertial confinement fusion (ICF) requires formation of a  high-temperature (>10 keV) central hot spot \citep{ICF1,ThermalCond}.  \cite{ICF1} found a turbulent kinetic energy cascade at the location of the hot-spot in the absence of the viscosity.  However,  viscous effects are significant and strongly damp small-scale velocity structures, with thermal conductivity enhancing the effect of the viscosity. \cite{ThermalCond} studied the RTI in the ICF context and found that the extremely high heat conduction and viscosity  can completely suppress the RTI development.  A detailed review on the effect of the viscosity is given in \cite{ZHOU1}.

The effects of the partial ionization on the development of the RTI have not been studied in detail and the results are mostly observational.  Observations have shown that ionization processes might occur at the interface between the two fluids. In the observations of the RTI in the solar atmosphere the acceleration is due to  gravity. Generally, in astrophysics   acceleration may also have other origins, such as the pressure of the confined relativistic plasma which accelerates a ``skin'' of ejecta at the boundary of the Crab Nebula \citep{prom2}.  The skin  is composed of   highly ionized material  and, like the inner filaments present inside the nebula, is brighter, most possibly because of the higher (electron) density that could be achieved by the ionization of the neutrals \citep{prom3,prom2}. 

In the solar context the RTI has been observed at the prominence corona transition region (PCTR), often at the border of prominence bubbles \citep{Berger_2017}. Small-scale upflows, also called plumes, have been  seen at the top border of the bubbles.  It is believed that these highly dynamic columns of plasma
appear due to the RTI \citep{2010Berger,2014Ozorco,Berger_2017}.  \cite{Berger_2017} observed a layer of  enhanced emission situated at the edges of the bubble{; as discussed below, these observations} can be explained as ionization effects. The PCTR is most susceptible to partial ionization effects because of the drop in collisional frequency towards the corona.

There are only a few analytical and numerical studies of the RTI in partially ionized solar plasmas, and most of them are carried out within the single-fluid approach. These studies typically use an MHD approach where the presence of the neutrals is introduced through the ambipolar term in the generalized Ohm's law
\citep{Arber_2007,sykora-rti,DiazKh2013,2014bKh}.  In an analytical study using a single-fluid model with ambipolar diffusion, \cite{DiazKh2013} find that decoupling between neutral and ionized plasma components can eliminate the stabilizing effect of the magnetic field for small scales.  The cutoff imposed by the component of the magnetic field parallel to the direction of the perturbation is removed when the ambipolar diffusion is included, and the RTI growth rate increases with neutral fraction. The result has been confirmed by numerical simulations of \cite{2014bKh}.  On the other hand, in a study using a two-fluid model, \cite{Diaz2012} find that the growth rate is reduced in the presence of partial ionization effects.  It is not yet fully clear what role partial ionization effects play in the development of RTI in prominences, as only a few theoretical studies have been performed to date. 

This study extends the work described in \cite{Popescu3}.  One of the conclusions of \cite{Popescu3} is that the linear growth of the RTI can be reduced due to damping introduced by collisional effects within and between ionized and neutral particle populations. The conclusion was made after comparing linear growth rates calculated from two-fluid simulations to growth rates computed semi-analytically within a single-fluid ideal MHD model. This  semi-analytical model was demonstrated to show higher growth rates than the simulations on smaller scales where collisional effects play significant role in the simulations.  

In this paper we systematically investigate the role of the collisional effects by turning off and on the viscosities and neutral thermal conductivity, the ionization--recombination processes, or by artificially increasing the ion-neutral elastic collision frequency within the two-fluid numerical model.  
{ We study
the impact of the collisional effects on the RTI growth rate, on its nonlinear
development and structure formation, and on potentially observable two-fluid
properties of RTI in solar prominences such as flow decoupling between charges
and neutrals }

We describe the setup and the simulations in Section~\ref{sec_setup}.  We study the effect of the inelastic collisions between neutrals and charges in Section~\ref{sec_rti2_inelastic}. The effects of { varying the collision frequency of} elastic collisions between particles that are different (charges and neutrals) and alike  (viscosity and thermal conductivity) {are discussed in Section~\ref{sec_rti2_elastic}, where an overview is followed by a study of the RTI mode content in Section~\ref{ssec:mode_analysis} and a detailed examination of ion-neutral flow decoupling in Section~\ref{ssec:decoupling}.}  The conclusions are drawn in Section~\ref{sec:conc}.

\section{Description of the problem}\label{sec_setup}
The simulations presented in this paper use the same 2.5D initial configuration {(i.e., physical quantities that vary in two dimensions, with vector variables with three components)} and governing two-fluid equations as described in Section~2 of \cite{Popescu3}.
\noindent
The prominence is represented by an enhancement in the neutral density with a corresponding decrease in the temperature, which is the same for neutrals and charges in equilibrium. { It is located at the center of the domain. The transition between the prominence and the coronal part of the domain is smooth, and the magnetic field can either be perpendicular to the perturbation plane or inclined and sheared over a distance $L_s$, over which it rotates by one degree to the plane.} In the simulations presented { below we} consider a shear length $L_s=L_0/2$, { where $L_0$ is a characteristic length scale of our domain, chosen to be equal to 1 Mm. The $L_s$ used here is smaller than the} value used in the simulations presented in \cite{Popescu3}, where $L_s=L_0$ was used.  As discussed in \cite{Popescu3}, shorter shear length accentuates the effects of magnetic field on RTI development, which we find to be helpful in comparing against simulations with uni-directional magnetic field background.  All simulations presented here are initialized with the multimode perturbation described by Eq.~11 in \cite{Popescu3}.

\begin{table*}[!ht]
\caption{List of the simulations. The columns, from left to right, are the short names of the simulations that appear in the text, the figures where they are referenced, and the parameters of the simulations}
\begin{tabular}{llp{11cm}}
\hline                  
Name & Figures & Parameters \\
\hline                  
P  & \ref{fig:comp-ct21}, \ref{fig:comp-ct22}, \ref{fig:time_snaps}, \ref{fig_group_growing}, \ref{fig:dec_snap} & Perpendicular magnetic field  described by Eq.~12 in \cite{Popescu3}.  
Ionization \& recombination terms turned on. Neutral viscosity \& thermal conductivity turned on. Multimode perturbation with $\delta=10^{-2}$ in Eq.~10 in \cite{Popescu3}.\\
P-I0  & \ref{fig:comp-ct21}, \ref{fig:comp-ct22} &  Same as P, but with ionization \& recombination terms turned off ($\Gamma_{\rm ion}=0$, $\Gamma_{\rm rec}=0$). \\
P-V0 & \ref{fig:time_snaps}, \ref{fig_group_growing}, \ref{fig:dec_snap} &  Same as P, but with neutral viscosity ($\xi=0$) \& thermal conductivity turned off ($K_n=0$). \\ 
L2  &  \ref{fig:sh-0.5-alpha1}, \ref{fig:sh-0.5-alpha2},  \ref{fig_group_growing}, \ref{fig:dec_snap} & Sheared magnetic field described by Eq.~13 in \cite{Popescu3} with $L_s=L_0/2$.  
Ionization \& recombination terms turned on. Neutral viscosity \& thermal conductivity turned on.
Multimode perturbation with $\delta=10^{-4}$ in  Eq.~10 in \cite{Popescu3}. \\
L2-$\alpha$ & \ref{fig:sh-0.5-alpha1}, \ref{fig:sh-0.5-alpha2}, \ref{fig_group_growing},  \ref{fig:dec_snap} & Same as L2, but with elastic collisions artificially (collisional parameter $\alpha$) increased by a factor of 10$^4$. \\
\hdashline \\
L2-S & \ref{fig:gr_k_elastic}, \ref{fig:velEigen} & Same as L2, but with a weaker multimode perturbation with $\delta=10^{-4}$ in Eq.~10 in \cite{Popescu3}.   \\
L2-$\alpha$-S  & \ref{fig:gr_k_elastic}, \ref{fig:velEigen} & Same as L2-$\alpha$, but with a weaker multimode perturbation with $\delta=10^{-4}$ in Eq.~10 in \cite{Popescu3}. \\  
L2-$\alpha$-V0-S & \ref{fig:gr_k_elastic} & Same as L2-$\alpha$, but with a weaker multimode perturbation with $\delta=10^{-4}$ in Eq.~10 in \cite{Popescu3} and with neutral viscosity \& thermal conductivity turned off.\\
L2-$\alpha$-V0-S,res/2 &  \ref{fig:gr_k_elastic}  & Same as L2-$\alpha$-V0-S, but with two times coarser resolution in both $x$ and $z$ directions. \\
\hdashline \\
${\rm P}_{\rm eq}$  & \ref{fig:comp-ct22},  \ref{fig:vel_snap}, \ref{fig:dec_snap}, \ref{fig:dec_hist} & No perturbation, only equilibrium is evolved. Perpendicular magnetic field  described by Eq.~12 in \cite{Popescu3}. Ionization \& recombination terms turned on. Neutral viscosity \& thermal conductivity turned on.\\
${\rm P\text{-}V0}_{\rm eq}$  & \ref{fig:comp-ct22},  \ref{fig:vel_snap}, \ref{fig:dec_snap}, \ref{fig:dec_hist} & Same as ${\rm P}_{\rm eq}$, but with neutral viscosity \& thermal conductivity turned off.\\
(P-I0)$_{\rm eq}$  & \ref{fig:comp-ct22} & Same as ${\rm P}_{\rm eq}$, but with ionization \& recombination terms turned off.\\
${\rm L2}_{\rm eq}$  &  \ref{fig:dec_hist}, \ref{fig:vel_snap}, \ref{fig:dec_snap} & No perturbation, only equilibrium is evolved. Sheared magnetic field described by Eq.~13 in \cite{Popescu3} with $L_s=L_0/2$. Ionization \& recombination terms turned on. Neutral viscosity \& and thermal conductivity turned on.   \\
${\rm L2\text{-}}\alpha_{\rm eq}$  &  \ref{fig:dec_hist},  \ref{fig:dec_snap}\ & Same as ${\rm L2}_{\rm eq}$, but with elastic collisions artificially increased by a factor of 10$^4$.\\
\hline
\end{tabular} \label{tab:param_tests2}
\end{table*}

Table~\ref{tab:param_tests2} shows a list of all simulations presented or referenced in this paper. 
 The following parameters  are changed throughout the
simulations:
\begin{enumerate}[\label{}]
\item {\it Magnetic field configuration.} We use either the uni-directional
perpendicular field configuration (simulation names that start with
P) or the sheared configuration (names start with
L2);
\item {\it Inelastic collisions.} We turn off ionization--recombination effects
in the governing two-fluid equations (names   with the suffix
-I0);
\item {\it Elastic collisions between like particles.}  We turn off the neutral
viscosity and thermal conductivity terms in the governing equations, which {corresponds to the limit of infinitely high collision frequency between like
particles} (names with the suffix -V0);
\item {\it Elastic collisions between charged particles and neutrals.}  We
increase the elastic collision frequency by multiplying the value of the
collisional parameter $\alpha$, calculated as in Eq.~A.4 in \cite{Popescu+etal2018},
by a factor of $10^4$  (names with the suffix   -$\alpha$);
\item {\it Amplitude of the perturbation.} We vary the amplitude of the perturbation
 through the parameter $\delta$ in Eq.~10 in \cite{Popescu3} using the value $\delta=10^{-2}$ in most of the simulations. However, for the
study of the linear growth rate, we   used an  amplitude smaller by a factor
of 100, i.e. $\delta=10^{-4}$ (names with  an additional suffix -S).
\end{enumerate}

The initial equilibrium is an ideal MHD equilibrium and it assumes complete collisional coupling. After introducing it into a two-fluid code, even without a perturbation associated with the RTI, the proper equilibrium evolves due to relative drift between charges and neutrals moderated by the elastic collisional terms, and due to non-equilibrium ionization and recombination (the ionization--recombination effect can be non-negligible). In order to separate, to the first order, the evolution of the background from the evolution due to RTI, we  performed simulations of the evolution of the equilibrium, indicated by the   suffix ``-eq'' in Table~\ref{tab:param_tests2} and in the text. 
In all the simulations we used the same boundary conditions and numerical procedure as described in \cite{Popescu3}.

\begin{figure}[!t]
 \includegraphics[width=8cm]{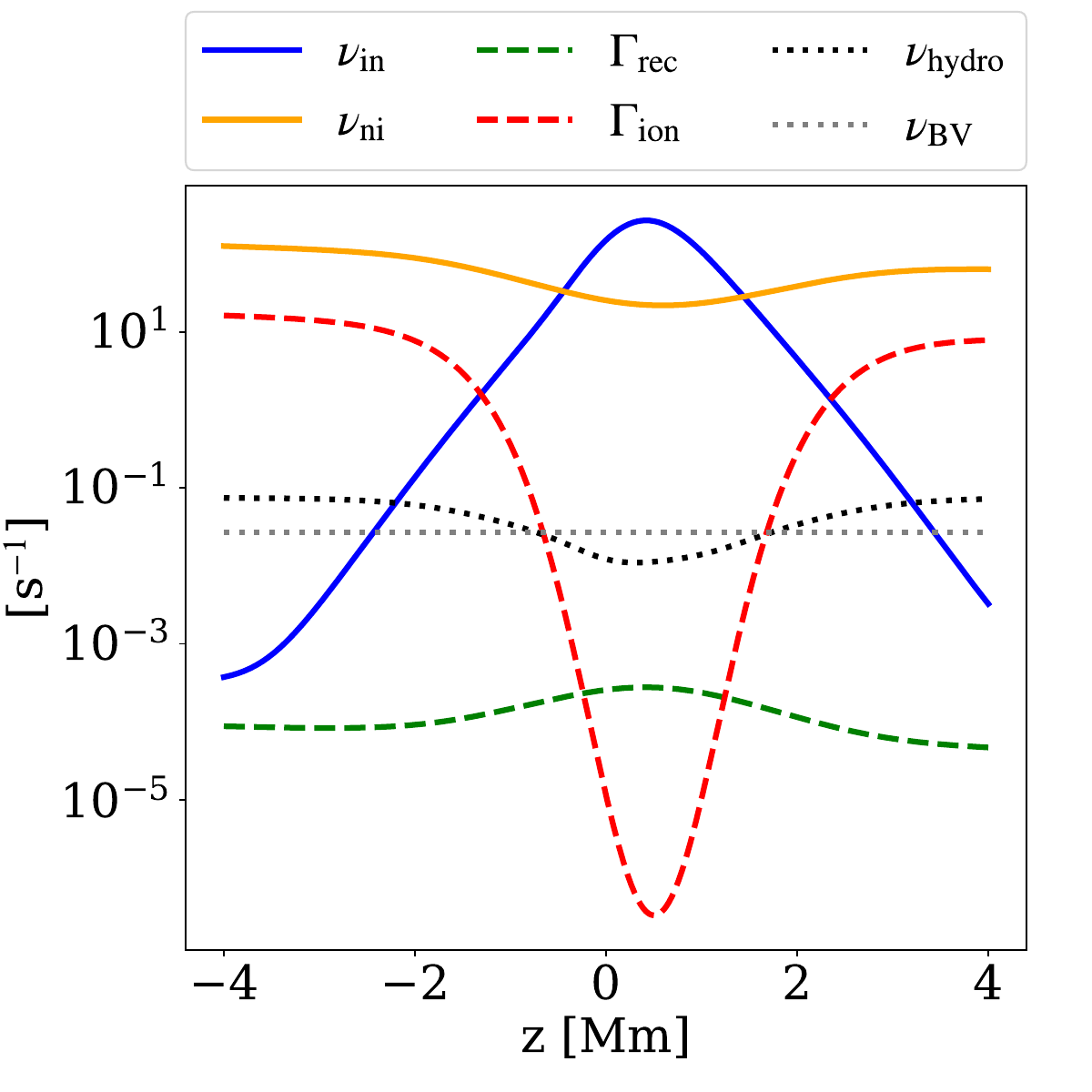}
\caption{Collision frequencies between ions and neutrals (solid blue line), neutrals and ions (solid orange line),  ionization rate (dashed red line) and recombination rate (dashed green line), and the hydrodynamic frequency (dotted black line) as a function of height, computed for the equilibrium model atmosphere. {For reference, the Brunt-V\"ais\"al\"a frequency  is calculated at height $z=0$ and is shown as a gray dotted line.}}
\label{diff_coef1}
\end{figure}

The equilibrium density and temperature height profiles for charges and neutrals assumed in the simulations lead to a particular height dependence of the collisional and characteristic hydrodynamic wave frequencies.  
{ Figure~\ref{diff_coef1} shows the ion-neutral, neutral-ion collision frequencies, ionization and recombination  rates, and the hydrodynamic frequency, as a function of height.}
The expressions for the ion-neutral and neutral-ion collision frequencies, and for the recombination--ionization rates are given in \cite{Popescu+etal2018} (Eqs. 5, A1, A2, {where we used $n_e = \rho_c /m_H$; $T_e = T_c$}).  The hydrodynamic frequency {is estimated} as $\nu_{\rm hydro}=c_0/L_0$, where $c_0$ is the sound speed of charges and neutrals together. { Another way to estimate the hydrodynamic timescale for the RTI is the characteristic Brunt-V\"ais\"al\"a frequency (dotted gray line).  The Brunt-V\"ais\"al\"a frequency is calculated as $\nu_{\rm BV} = \sqrt{g/L_d}$, where $g=273.98$~m/s$^2$ is the gravitational acceleration and $L_d = \rho_0/\frac{\partial \rho_0}{\partial z}$ is the gradient scale length of total density at $z=0$. The two hydrodynamic timescale estimates give similar values, as shown in Figure~\ref{diff_coef1}.}  

We observe that there are regions where the timescales associated with ion-neutral interactions are larger (their frequencies are lower) or smaller (their frequencies are higher) than those of the hydrodynamic evolution. The recombination (blue solid line) evolves on a timescale that is three orders of magnitude slower than hydrodynamic evolution, and thus is unlikely to impact it. The neutral-ion collisions evolve on a  much faster timescale than the hydrodynamic timescale, implying that the neutrals are closely coupled to ions by elastic collisions. On the other hand, the ion-neutral collision frequency and the ionization rate become comparable to the hydrodynamic frequency in some regions of the domain.  The ion-neutral collision frequency is higher in the middle of the domain and becomes lower towards the edges (reducing the collisional coupling), while the ionization rate becomes higher at the edges. Therefore, ionization and ion-neutral coupling are expected to directly affect the hydrodynamic evolution, highlighting the need for a two-fluid model where the charges and neutrals are not assumed to be fully coupled.  In contrast, any single-fluid model attempting to approximate partial ionization effects has to assume instantaneous and functionally prescribed changes to the local degree of ionization, which may omit critical nonlinear two-fluid effects (see, e.g.,  \citealt{2013Lukin,Ni2018a,Ni2018b}).


\section{Effect of  inelastic collisions} \label{sec_rti2_inelastic}

\begin{figure}[!t]
 \centering
 \includegraphics[width=8cm]{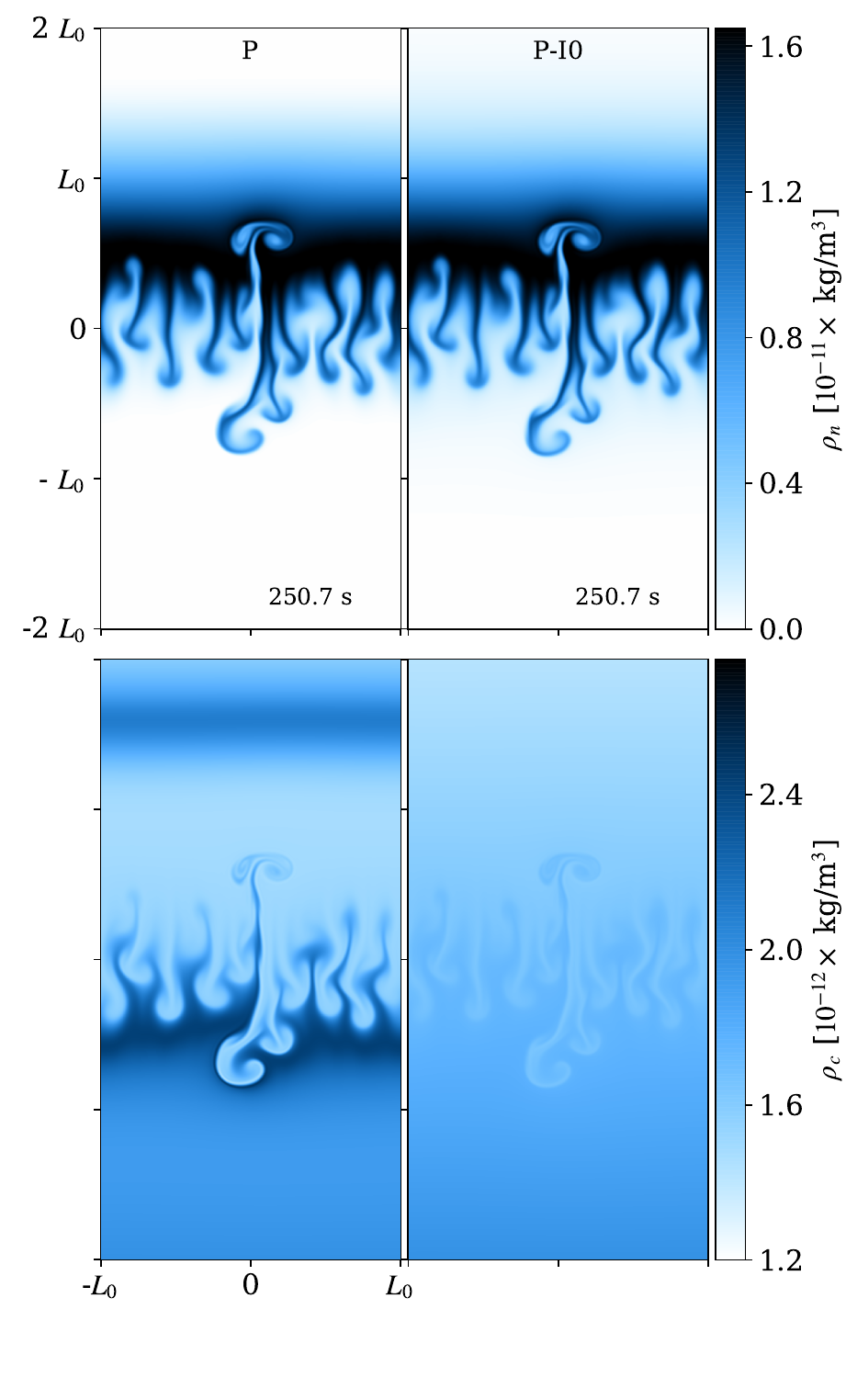}
\caption{{ Snapshots of the neutral density (top) and charge density (bottom) for the simulations P (left) and P-I0 (right). The snapshots were taken at the same moment in time. Here and below, only a portion of the domain between -2 $L_0$ and 2 $L_0$ in the $z$ direction is shown.}}
\label{fig:comp-ct21}
\end{figure}

\begin{figure}[!ht]
\centering
\includegraphics[width=8cm]{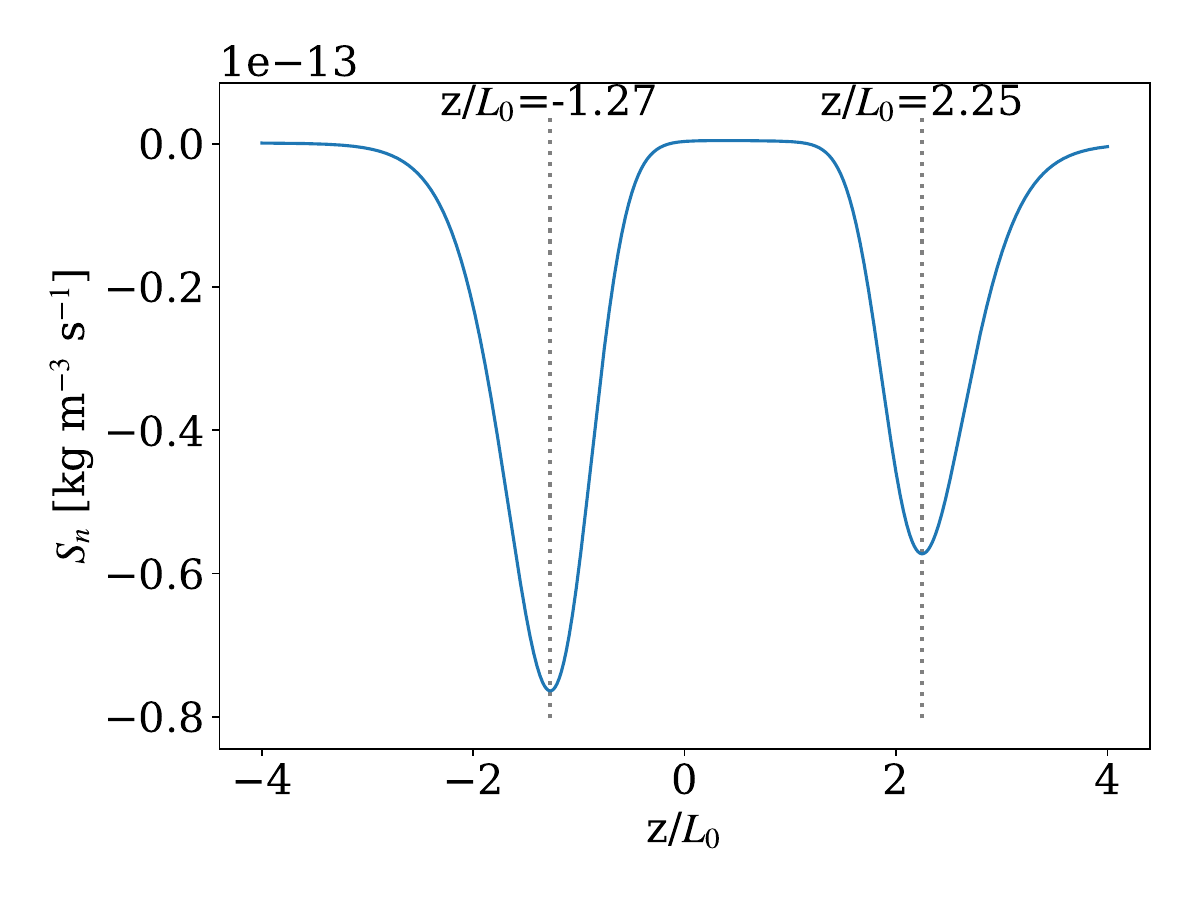}
\caption{Source term from the neutral continuity equations, $S_n$, calculated for the equilibrium atmosphere from Eq. \ref{eq:sn_equi} as a function of height.}
\label{fig:sn_inelastic}
\end{figure}

\begin{figure}[!ht]
 \centering
 \includegraphics[width=7.5cm]{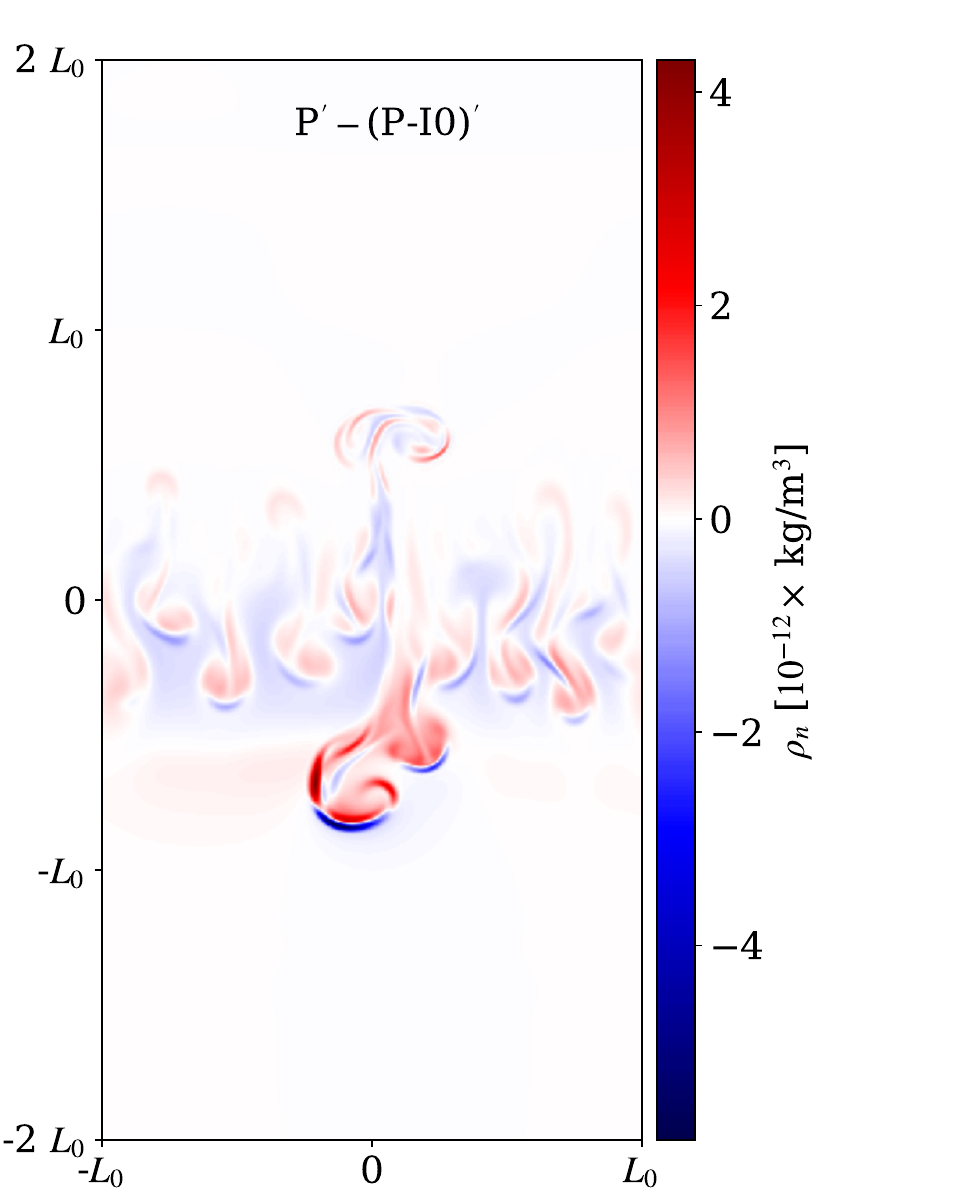}
\caption{Difference in the density of neutrals between simulations P and P-I0 at $t=250.7$~s.  For each simulation the density of neutrals obtained after evolving the unperturbed  equilibrium has been subtracted: ${\rm P}' \equiv {\rm P} - {\rm P}_{\rm eq}$; $\text{\rm (P-I0)}' \equiv \text{\rm P-I0} - \text{\rm (P-I0)}_{\rm eq}$.}
\label{fig:comp-ct22}
\end{figure}


To study the possible effects of inelastic collisions, such as electron impact ionization, on the development of RTI we compare simulations with and without ionization--recombination terms in the governing equations. Here we focus on simulations with a uni-directional background magnetic field that is perpendicular to the plane of the { perturbation}.

A comparison of linear RTI growth rates in simulations with and without ionization and recombination { (not shown here) } reveals virtually no difference between the two for any of the linearly unstable modes. 
However, nonlinear evolution begins to show some differentiation between the simulations.
Figure~\ref{fig:comp-ct21} { shows the neutral and charged densities in the }simulations P and P-I0 in the nonlinear phase at $t = 250.7$~s.  We observe no easily discernible difference between the two snapshots of neutral density, but there is an increased contrast in the density of charges in the P simulation  compared to P-I0. The sharp edge of the structures developed in the charged fluid must be due to the ionization of the edge of the prominence thread in the hot surrounding corona. 

We note that the initial background atmosphere in the simulations is not in perfect ionization--recombination equilibrium and evolves slowly.  In Figure~\ref{fig:sn_inelastic} we show the source term in the continuity equation of the neutral species calculated with equilibrium variables, 
\begin{equation}\label{eq:sn_equi}
S_{\rm n} = \rho_{\rm c0} \Gamma^{\rm rec} - \rho_{\rm n0} \Gamma^{\rm ion}\,,
\end{equation}
as a function of height.  We observe that this quantity is not zero everywhere, and has two minima located at $z=-1.27$~Mm and $z=2.25$~Mm, which lead to a gradual increase in the density of charges at the edges of the prominence thread that can be observed in all simulations with ionization--recombination, in particular  for  simulation P in Figure~\ref{fig:comp-ct21}.  We also note that even in the absence of this artifact of the initial condition, due to imperfect ion-neutral coupling, the neutral density concentration associated with a weakly ionized prominence thread invariably spreads vertically due to $\nabla p_n$ forces, and drifts downward due to gravity.  
{ Thus, it is reasonable to expect that any long-lived prominence thread may gradually evolve, in part, due to ionization at the edges of the thread.}

To better illustrate the  ionization--recombination effects on nonlinear evolution of RTI, Figure~\ref{fig:comp-ct22} shows the difference in the neutral density structures between the P and P-I0 simulations after subtracting the respective evolution of the unperturbed equilibrium for each simulation. We observe that the magnitude of the maximum difference between the two simulations is $\approx 4\times10^{-12}~\text{kg/m}^3$, or $\approx 25\%\text{-}30\%$ of the peak neutral mass density values and greater than the maximum mass density of charges shown in Fig.~\ref{fig:comp-ct21}.  The relative excess and depletion of neutral mass density are observed to have approximately the same magnitude and are located near the main spike of falling weakly ionized material, indicating that the prominence material falls slightly faster in the simulation that neglects ionization--recombination effects.  We measure the difference between the two simulations in the downward velocity of the spike at this time to be $\approx2\text{-}3$~km/s, with the prominence material falling at $\approx12\text{-}15$~km/s (not shown).  While it is a subtle effect this early in the formation and propagation of the spike, the difference in the evolution can be expected to grow with time leading to overestimation of the rate of the downward spike propagation in simulations that neglect ionization--recombination effects.

%
\section{{Effect of   elastic collisions}} \label{sec_rti2_elastic}
%

\begin{figure}[!ht]
     \includegraphics[width=8cm]{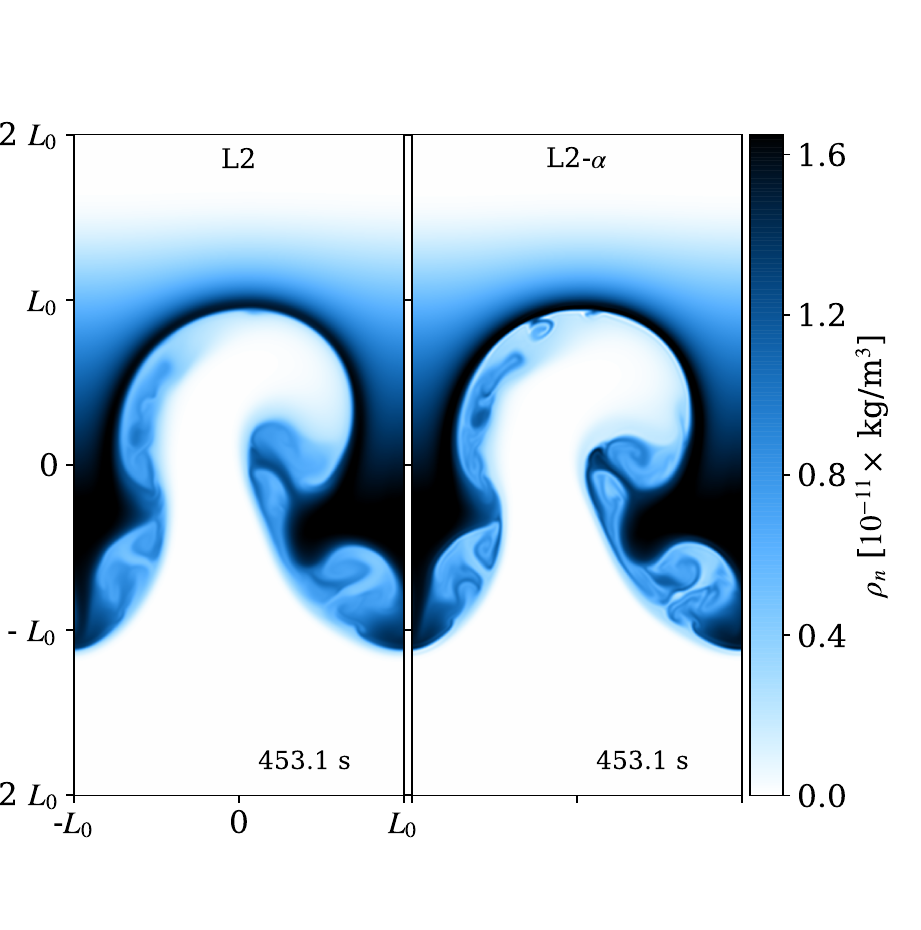}
\caption{
Comparison of the snapshots of the density of neutrals  for the simulations with sheared magnetic field with the self-consistent ion-neutral collision frequency, L2 (left panel), and with artificially increased ion-neutral collision frequency, L2-$\alpha$ (right panel).}
\label{fig:sh-0.5-alpha1}
\end{figure}
\begin{figure}[!ht]
 \includegraphics[width=8cm]{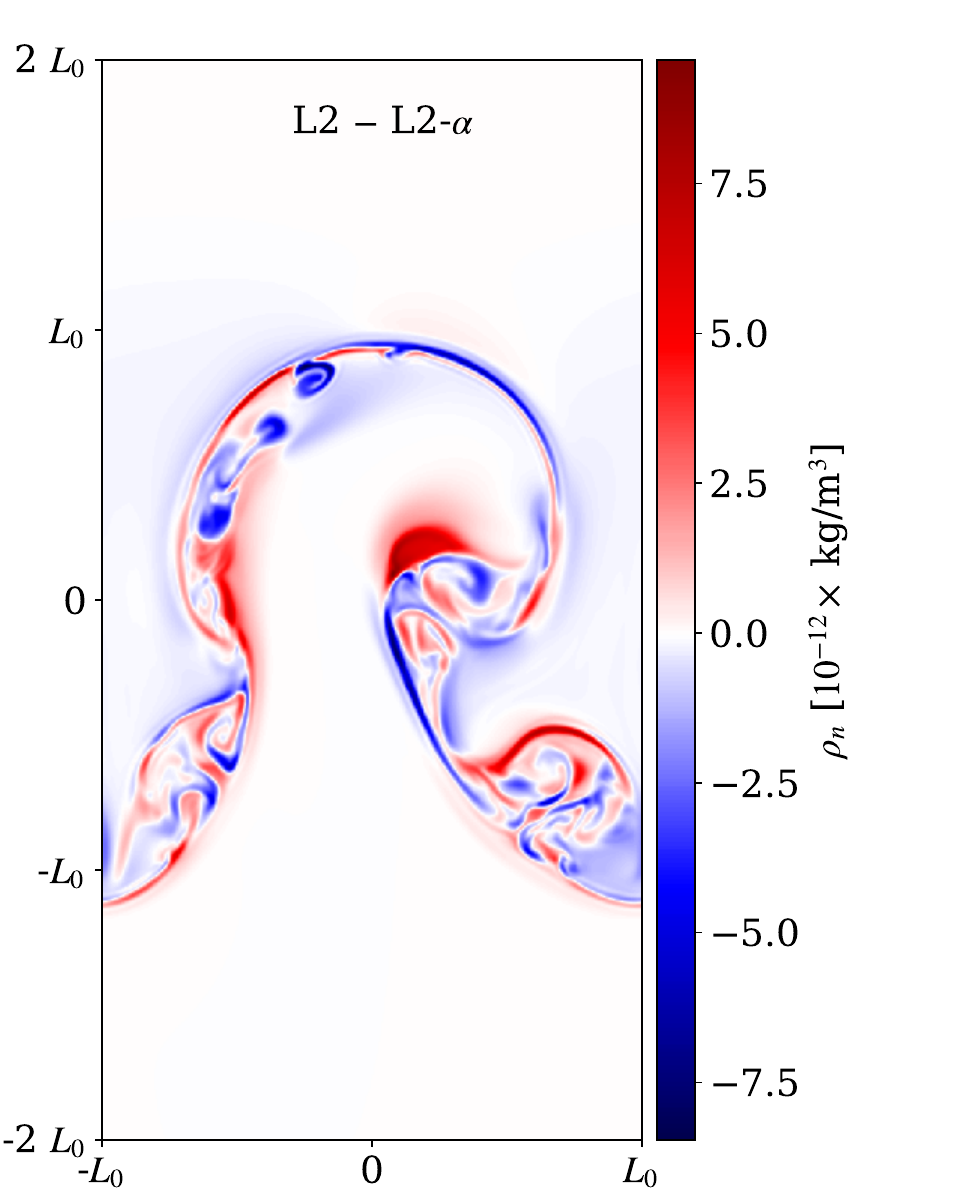}
\caption{Difference in neutral densities between   simulation L2 and L2-$\alpha$ at $t=453.1$~s.}
\label{fig:sh-0.5-alpha2}
\end{figure}

As discussed in \cite{Popescu3}, the solar prominence plasma conditions modeled in this study are such that RTI damping due to finite neutral-ion collision mean free path (MFP) begins to act on scales longer than those impacted by viscous damping. Furthermore, in sheared magnetic field configurations, the decoupling between charged and neutral species is exacerbated by the presence of a magnetic field component in the plane of plasma motion, which also acts to stabilize RTI modes on scales shorter than the prominence density gradient scale. On the other hand, the uni-directional magnetic field perpendicular to the plane of motion has no stabilizing effect on RTI at any scale and has little impact on the degree of decoupling between charged and neutral fluid motions.
Thus, to systematically explore the impacts of elastic collisions   between and within the particle species on RTI, we focus on the sheared magnetic field configuration, L2, for the former, and on the uni-directional perpendicular magnetic field configuration, P, for the latter.

We first illustrate the impact of the finite MFP between charged particles and neutrals. To do so, we compare simulations L2 and L2-$\alpha$, run under the same conditions,  the only difference being the collisional frequency between charged particles and neutrals.  The L2-$\alpha$ simulation has both $\nu_{in}$ and $\nu_{ni}$ increased artificially by a factor of $10^4$, thus correspondingly reducing the MFP.  Figure~\ref{fig:sh-0.5-alpha1} shows snapshots of the neutral mass density from L2 and L2-$\alpha$ simulations at $t=453.1$~s.
It is apparent from the maps of the neutral density that { a greater number of smaller scales, and  density gradients that are even sharper, develop in the L2-$\alpha$ case}.


Figure~\ref{fig:sh-0.5-alpha2} quantifies the difference between the neutral density maps of the two simulations.  We observe that the maximum differences are  $\approx 50\%$ of the peak value of the neutral density in the prominence thread. The differences appear as numerous small-scale structures, consistent with the ion-neutral decoupling impacting small-scale density structure formation during nonlinear RTI development.

\begin{figure}[!t]
 \centering
 \includegraphics[width=8cm]{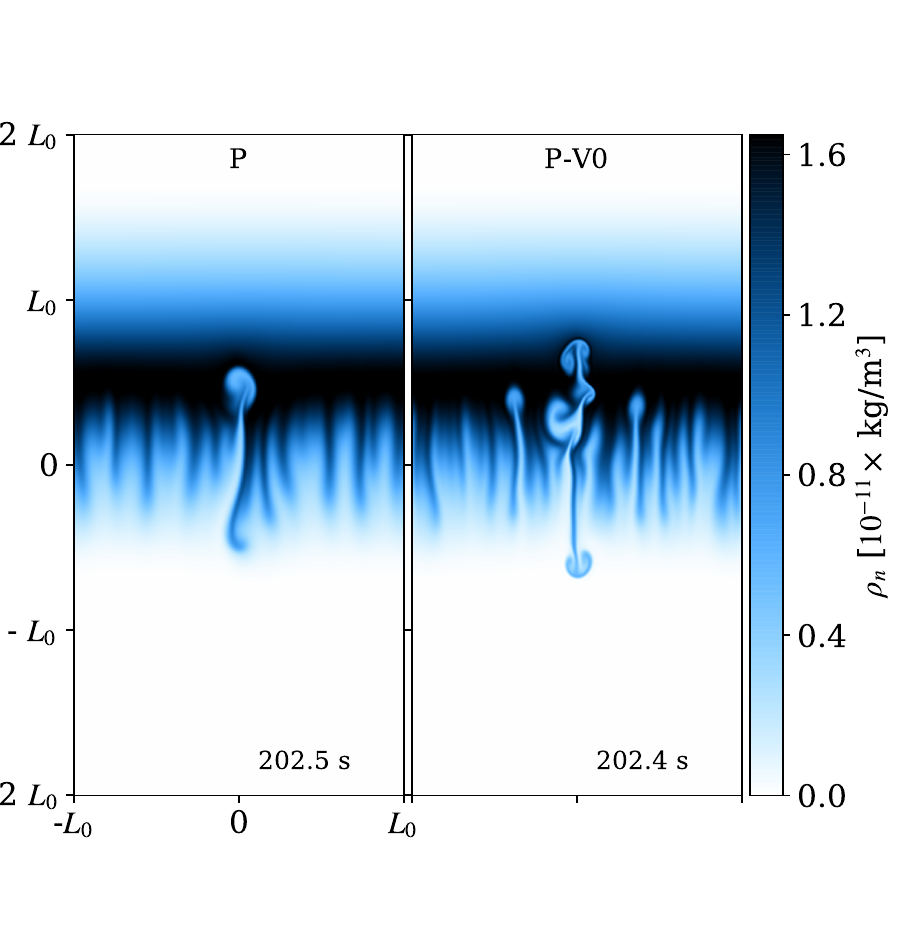}
\caption{Comparison of the snapshots of the density of neutrals for the simulations with perpendicular magnetic field P (left panel) and P-V0 (right panel).}
\label{fig:time_snaps}
\end{figure}

\begin{figure*}[!htbp]
 \centering
 \includegraphics[width=16cm]{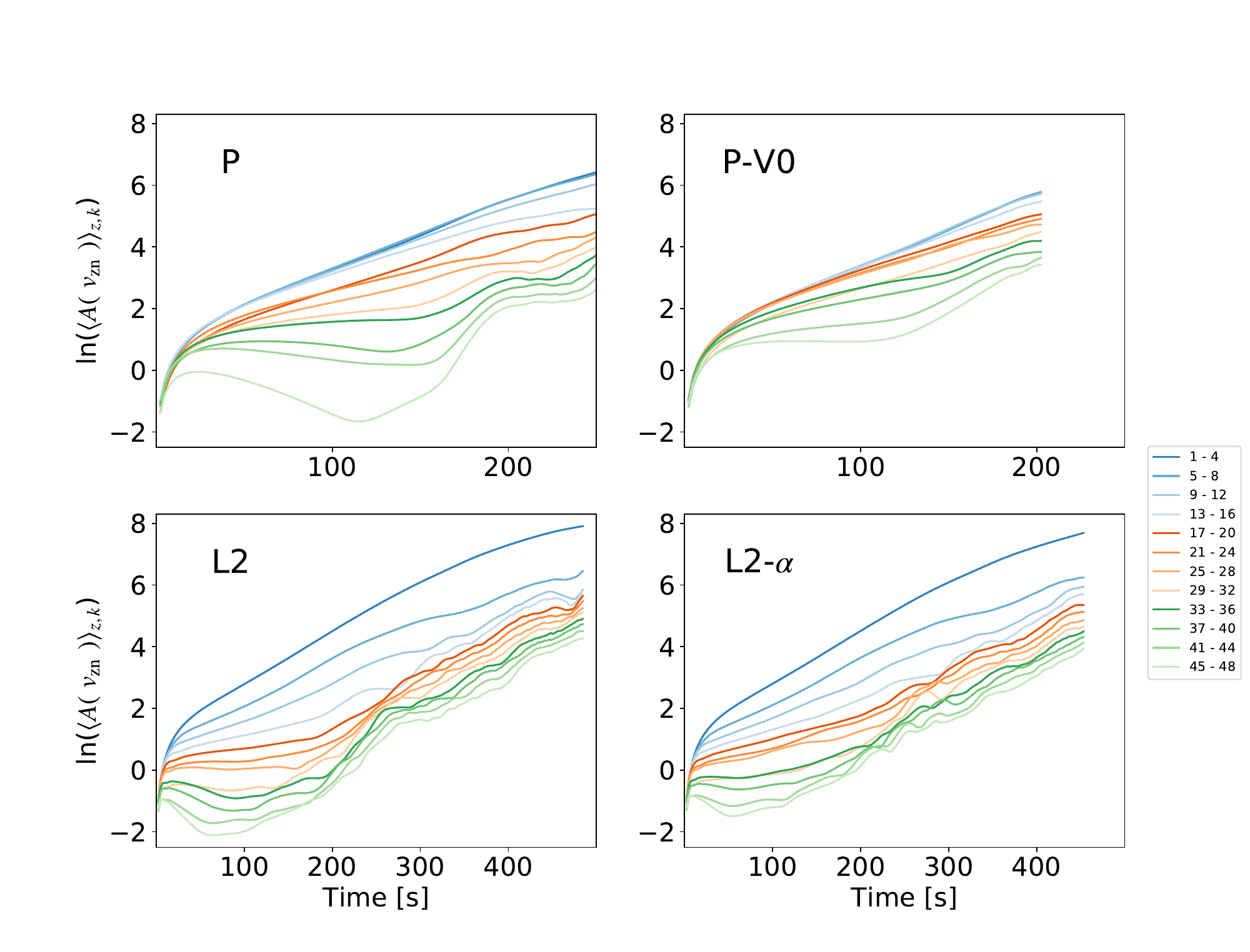}
\caption{Time evolution of individual horizontal mode Fourier harmonics, computed from the vertical velocity of neutrals, as a function of time. The  colored lines show the Fourier amplitudes of the harmonics, grouped in sets of four, from $n=1$ to $n=48$ (see legend at right). The amplitudes are averaged between heights -$L_0$ and $L_0$ and within each group. The results for four simulations are shown: P (top left); P-V0 (top right); L2 (bottom left); L2-$\alpha$ (bottom right).
}
\label{fig_group_growing}
\end{figure*}

\begin{figure}[!htb]
 \includegraphics[width=8.5cm]{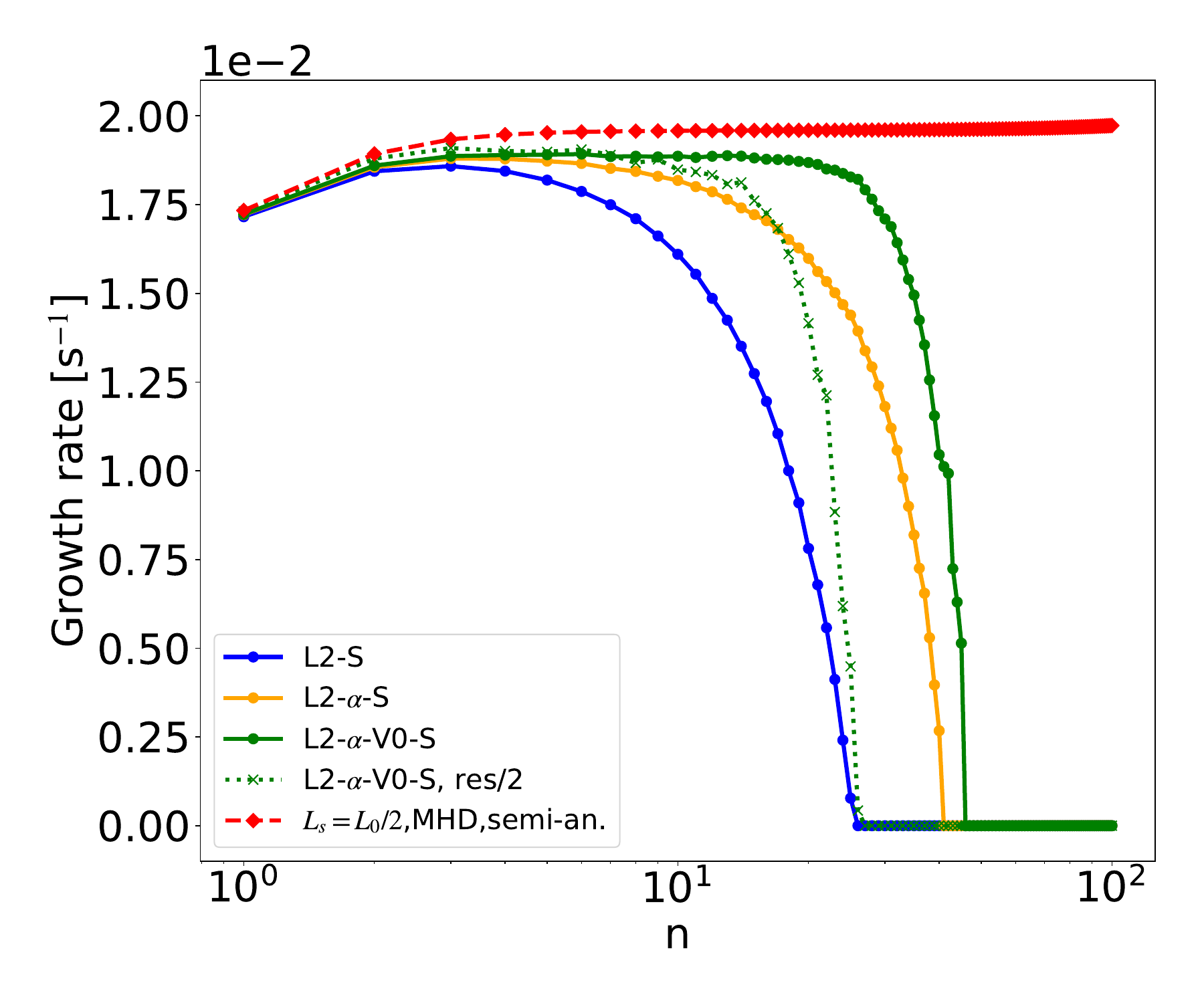}
\caption{Growth rate for the simulations L2-S (blue line), L2-$\alpha$-S (orange line), { L2-$\alpha$-V0-S} (green solid line), and { L2-$\alpha$-V0-S,res/2} for half resolution  in the vertical and the horizontal direction (green dotted line) as a function of the {horizontal} mode number n.
The red line shows the  semi-analytical solution
in the ideal incompressible MHD approximation.
}
\label{fig:gr_k_elastic}
\end{figure}
\begin{figure*}[!htb]
 \includegraphics[width=16cm]{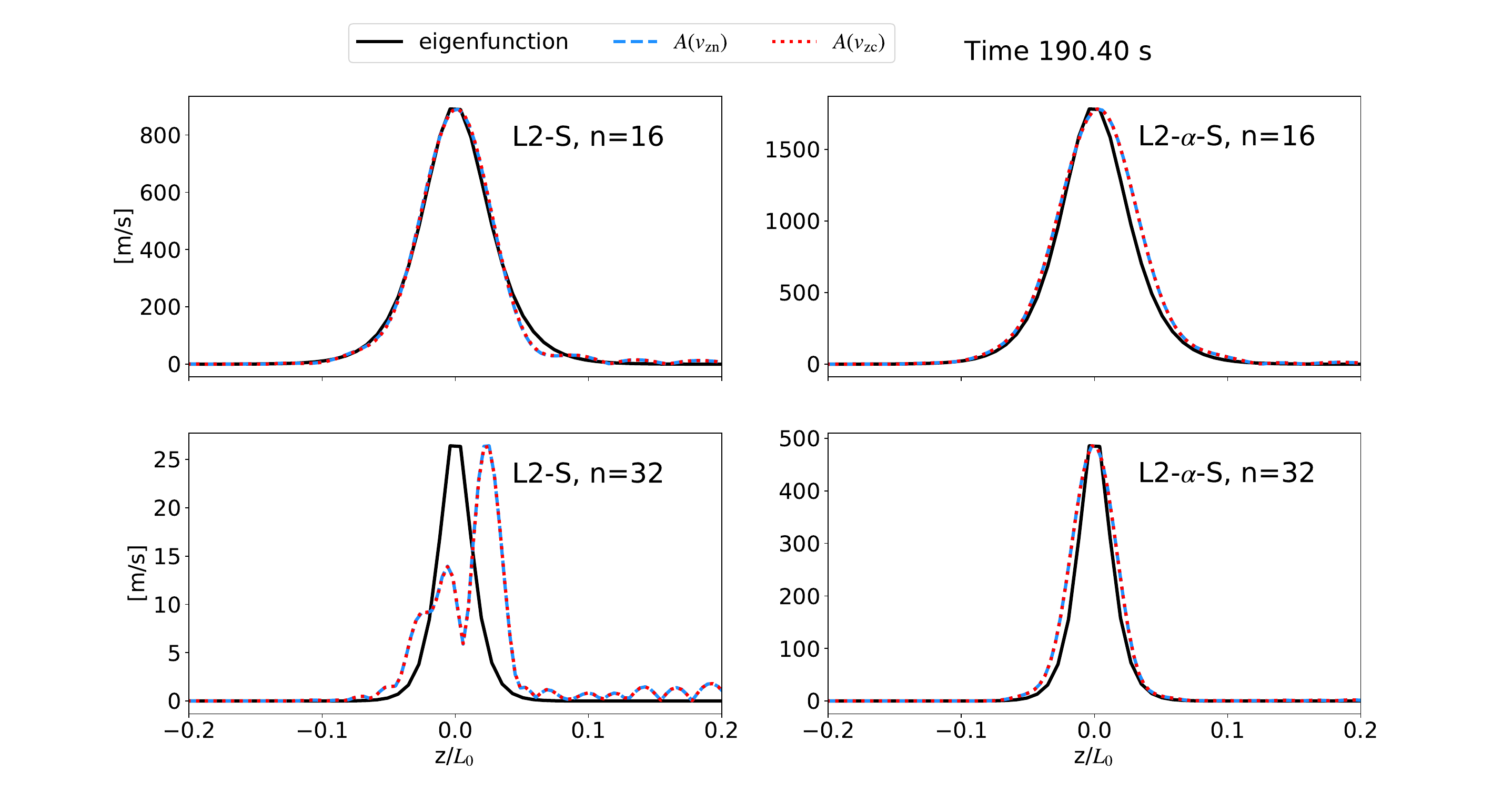}
\caption{Fourier amplitudes of the vertical velocities of neutrals (blue dashed line) and charges (red dotted line) at the end of the linear phase at time $\approx$ 190.4~s,
and the vertical eigenfunction of the velocity obtained in the { incompressible MHD approximation} (black solid line), as a function of height between -0.2 Mm and 0.2 Mm.
{\bf Top left panel:} L2-S, n=16;
{\bf Top right panel:} L2-$\alpha$-S, n=16;
{\bf Bottom left panel:} L2-S, n=32;
{\bf Bottom right panel:} L2-$\alpha$-S, n=32.
The eigenfunctions are normalized to the corresponding maximum of the velocity amplitudes. { The profile of the semi-analytical eigenfunction is less smooth as it was   obtained on a coarser grid of 1024 points, compared to simulations that used 2048 points in the vertical direction.}}
\label{fig:velEigen}
\end{figure*}
We next illustrate the effects of the viscosity and neutral thermal conductivity on RTI development.  To do so, we compare simulations P and P-V0, also run under the same conditions with the only difference being the lack of the viscous and heat conduction terms in the governing equations for the P-V0 simulation. Figure~\ref{fig:time_snaps} shows snapshots of the density of neutrals 
taken at nearly the same time in the two simulations.
We again observe that more developed smaller scales and sharper density gradients are present in the simulation that omits the diffusion effects, here associated with viscosity and thermal conductivity of neutrals.  We also note that the limit of zero viscosity corresponds to the limit of infinitely high collision frequency, and thus infinitely small collision MFP within the particle species.   

Figures~\ref{fig:sh-0.5-alpha1}-\ref{fig:time_snaps} clearly demonstrate that a finite MFP for elastic collisions in all cases leads to the formation of fewer small-scale density structures.  While this  conclusion could have been expected { from general considerations}, we now explore   the linear and nonlinear evolution of RTI in these simulations, as well as the consequences for potentially observable decoupling in flow between the species, in more detail.


\subsection{Mode analysis}\label{ssec:mode_analysis}

We analyzed the effect of the elastic collisions on the development of different spatial scales in the simulations.  Figure~\ref{fig_group_growing} shows the evolution of the amplitudes of horizontal Fourier modes of the vertical velocity of neutrals for the first 48 modes.  The Fourier amplitudes have been averaged in height between $-L_0$ and $L_0$, which is the region where the instability develops, and grouped and averaged in sets of four modes for better visualization { (for details, see \citealt{Popescu3})}.

The evolution of the RTI can be considered in two phases: in the first linear phase the amplitudes are small and modes are dominantly non-interacting; in  the second nonlinear phase the mode evolution is dominated by mode-mode interactions. %
It is apparent that the growth rate of large-scale modes, up to $n \approx 16$, is nearly identical between the cases within each pair of simulations considered: P and P-V0, and L2 and L2-$\alpha$. 
We also observe that the high-$n$ modes, { $n \gtrapprox 37$}, are suppressed during the early development of the instability and do not have a linear growth phase.  In the L2 simulations the growth of intermediate scales is also stabilized by the in-plane component of the magnetic field.  However, in both simulations where the collisional MFP was artificially decreased, the degree of damping of high-$n$ modes is visibly smaller {or altogether absent}.   

The growth of velocity amplitudes of the low-$n$ modes decreases at the beginning of the nonlinear phase. Through nonlinear mode-mode coupling, energy begins to be transferred to smaller scales, with the amplitude of high-$n$ modes rapidly increasing.  Here, the impact of finite elastic collision MFP is again apparent in comparing the magnitude of high-$n$ modes for P and P-V0 simulations; for example, at $t=200$~s the P-V0 simulation shows high-$n$ modes to be over an order of magnitude stronger than the P simulation.  There is less  difference between the L2 and L2-$\alpha$ simulations. This is indicative of the important role magnetic field dynamics plays in nonlinear RTI evolution within sheared field configurations \citep{Popescu3}, where Lorentz forces can drive small-scale flows against any damping introduced by finite collision MFP effects. 

{ A quantitative analysis of the effect of the collisions on the growth of RTI can be performed by calculating the growth rate for different modes. The linear growth rate calculated from the simulations can be compared to the analytical growth rate obtained within the ideal single-fluid MHD approximation where the damping effect of the collisions is neglected. We consider  the effect of the finite MFP for ion-neutral collisions on the growth rate for the sheared field case, where this effect is combined with the magnetic field stabilization effects (see below).  We then study the effect of the like particle collisions by further neglecting the viscosity and thermal conductivity, thus also allowing us to estimate the effects of numerical dissipation.}


Figure~\ref{fig:gr_k_elastic} shows the RTI linear growth rates as a function of mode number for four simulations: L2-S, L2-$\alpha$-S, L2-$\alpha$-V0-S, and L2-$\alpha$-V0-S,res/2. The growth rate from the simulations is calculated as in   \cite{Popescu3} by linear fitting the dependence of the individual mode amplitudes on time in the linear phase of the RTI. 
The errors of the linear fit, defined as the ratio of the covariance of the fitted coefficient to its value, are below 5$\times$10$^{-4}$ for all the modes shown in the plot. The semi-analytical growth rate (calculated as  described in \citealt{Popescu3} under the incompressible, ideal MHD assumption) is plotted as a reference (red line). 

We observe in Figure~\ref{fig:gr_k_elastic} that for the largest scales (n=1), the growth rate is the same for the four simulations, and is nearly identical to the semi-analytical calculation. The effect of the ion-neutral collisions becomes noticeable when the only simulation of the four that accounts for finite ion-neutral MFP effects (L2-S,  blue line) diverges from the others for $n \gtrapprox 5$, showing lower growth rate for higher-$n$ modes. The additional effect of viscosity is demonstrated by comparing results of the simulation with finite neutral-neutral collision MFP (L2-$\alpha$-S, orange line), and the simulation with zeroed-out viscous and heat conduction terms (L2-$\alpha$-V0-S,  solid green line). These diverge for modes with $n \gtrapprox 10$, confirming our previous estimate that for our atmosphere the finite ion-neutral collision MFP affects larger scales than the finite neutral-neutral collision MFP.  

The main difference between the semi-analytical calculation shown in red ($L_s = L_0/2$,MHD,semi-an.) and the L2-$\alpha$-V0-S simulation result, namely the presence of a cutoff at $n\approx 45$ in the simulation, should be attributed to numerical dissipation. This is confirmed by the fact that the cutoff mode number drops by a factor of almost two when the simulation resolution is decreased by a factor of two (green dotted line, labeled ``L2-$\alpha$-V0-S,res/2''). Importantly, it also implies that for all simulations analyzed elsewhere in this paper (i.e., those listed in the top and bottom portions of Table~\ref{tab:param_tests2}), the cutoff due to physical effects falls at lower $n$ than the cutoff produced by the numerical dissipation, and the modes affected by the physical dissipation can be expected to be well resolved.

Another way to quantify the effects of ion-neutral decoupling is by comparing the $z$-dependence of the RTI eigenmode amplitudes calculated semi-analytically in the single fluid MHD limit to the corresponding modes observed to grow in the simulations.  We use vertical velocity modes for this comparison.  In the semi-analytical calculation, the vertical profile of the velocity is obtained by calculating the eigenfunction $v_z(z)$ from the second-order ordinary differential equation, Eq.~(18) in \cite{Popescu3}.
In the simulations, the vertical profile of the growing modes is determined by taking a Fourier transform in the $x$-direction of the vertical velocities of neutrals and charges at each height at one time during the linear phase of the simulation.  

Figure~\ref{fig:velEigen} shows the Fourier amplitudes of $v_{zn}$ and $v_{zc}$ modes from simulations L2-S and L2-$\alpha$-S compared to the semi-analytically calculated $v_z$ eigenfunctions at heights between $z=-0.2$~Mm and $z=0.2$~Mm. The Fourier amplitudes are computed from snapshots taken at the end of the linear phase at $t=190.4$~s.  The semi-analytical eigenfunctions are normalized so that the maxima of their amplitudes coincide with those from the simulations.  Functional forms for two modes, $n=16$ (top row) and $n=32$ (bottom row), and for the two simulations, L2-S (left column) and L2-$\alpha$-S (right column), are shown. 

We observe that the RTI modes are narrower for higher mode numbers, as previously discussed by \cite{Popescu3}. This property makes the high-n modes more affected by the ion-neutral decoupling.  We also observe that the profile of the Fourier amplitudes of the velocities and the eigenfunction match each other closely for $n=16$ for both simulations (top row). However, this is not the case for the $n=32$ mode. The bottom right panel shows that even in the presence of finite neutral-neutral MFP effects, but with negligible ion-neutral collision MFP, the $n=32$ modes in the L2-$\alpha$-S simulation and the semi-analytical calculation are still closely matched. On the other hand, the same comparison for the L2-S simulation that includes finite ion-neutral MFP effects shows a qualitatively different mode shape.  The $n=32$ mode in the L2-S simulation also has a factor of $\approx 20$ lower amplitude than the same in the L2-$\alpha$-S simulation, even though the functional forms for $v_{zn}$ and $v_{zc}$ are indistinguishable from each other in the two simulations.
These comparisons again demonstrate the importance of accurately accounting for the finite ion-neutral MFP effects when modeling RTI in solar prominences.



%
\subsection{Decoupling} \label{ssec:decoupling}

\begin{figure}[!htb]
 \centering
 \includegraphics[width=8cm]{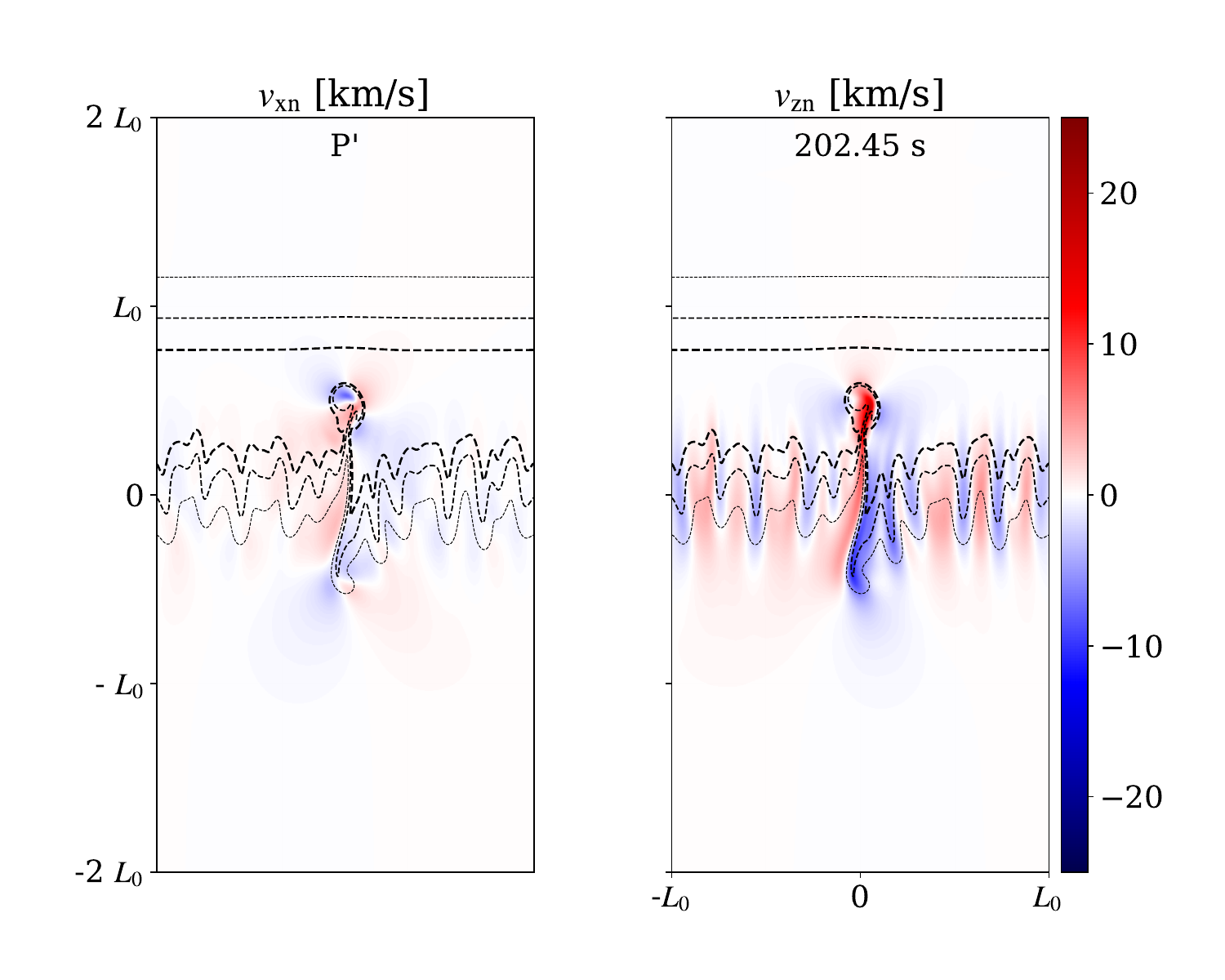}

 \includegraphics[width=8cm]{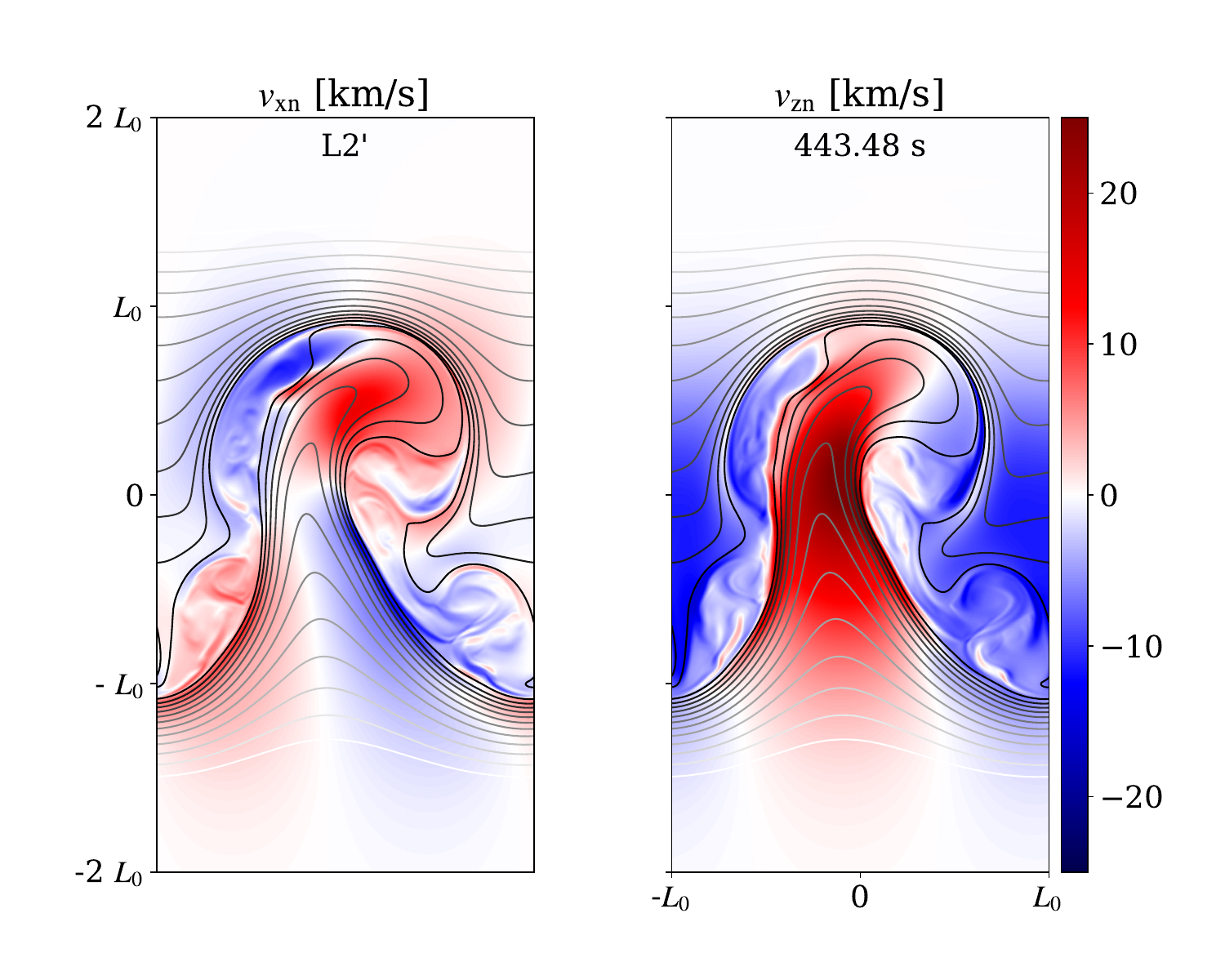}
\caption{{ Snapshots of the  horizontal (left) and vertical (right) velocities for the ${\rm P}' \equiv {\rm P} - {\rm P}_{\rm eq}$ case (top) and ${\rm L2}' \equiv {\rm L2} - {\rm L2}_{\rm eq}$  case (bottom), where the evolution of the background has been removed. The isocontours of the neutral density spanning the range from the minimum to the maximum value are plotted for the ${\rm P}'$ snapshots (top). The magnetic field lines are plotted over the ${\rm L2}'$ snapshots as isocontours of the magnetic potential (bottom), }with levels corresponding to values from 0.6$A_{\rm y}^{\rm max}$ and $A_{\rm y}^{\rm max}$, where $A_{\rm y}^{\rm max}$ is the maximum value of $|A_{\rm y}|$.}
\label{fig:vel_snap}
\end{figure}

\begin{figure*}[!htb]
 \centering
 \includegraphics[width=8cm]{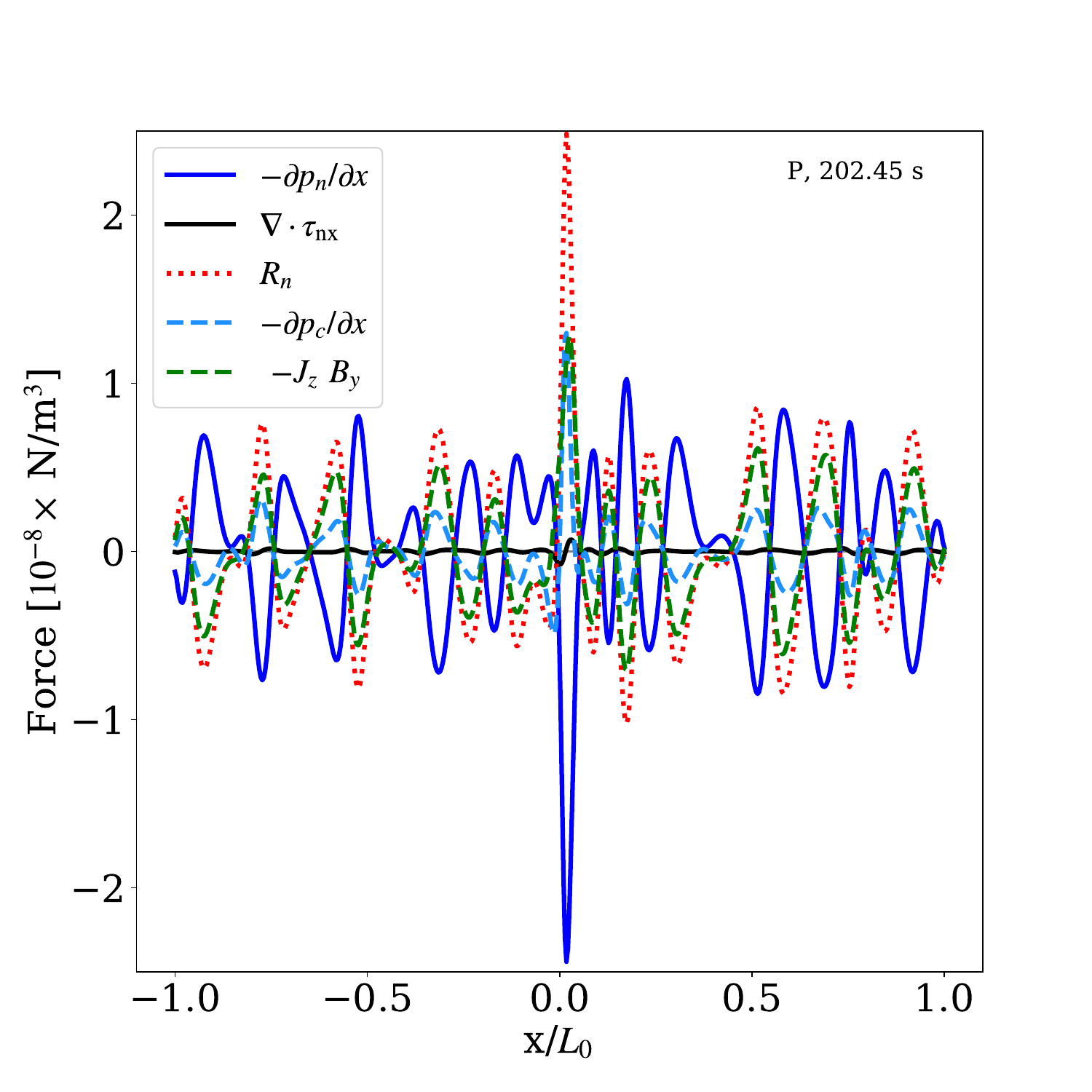}
 \includegraphics[width=8cm]{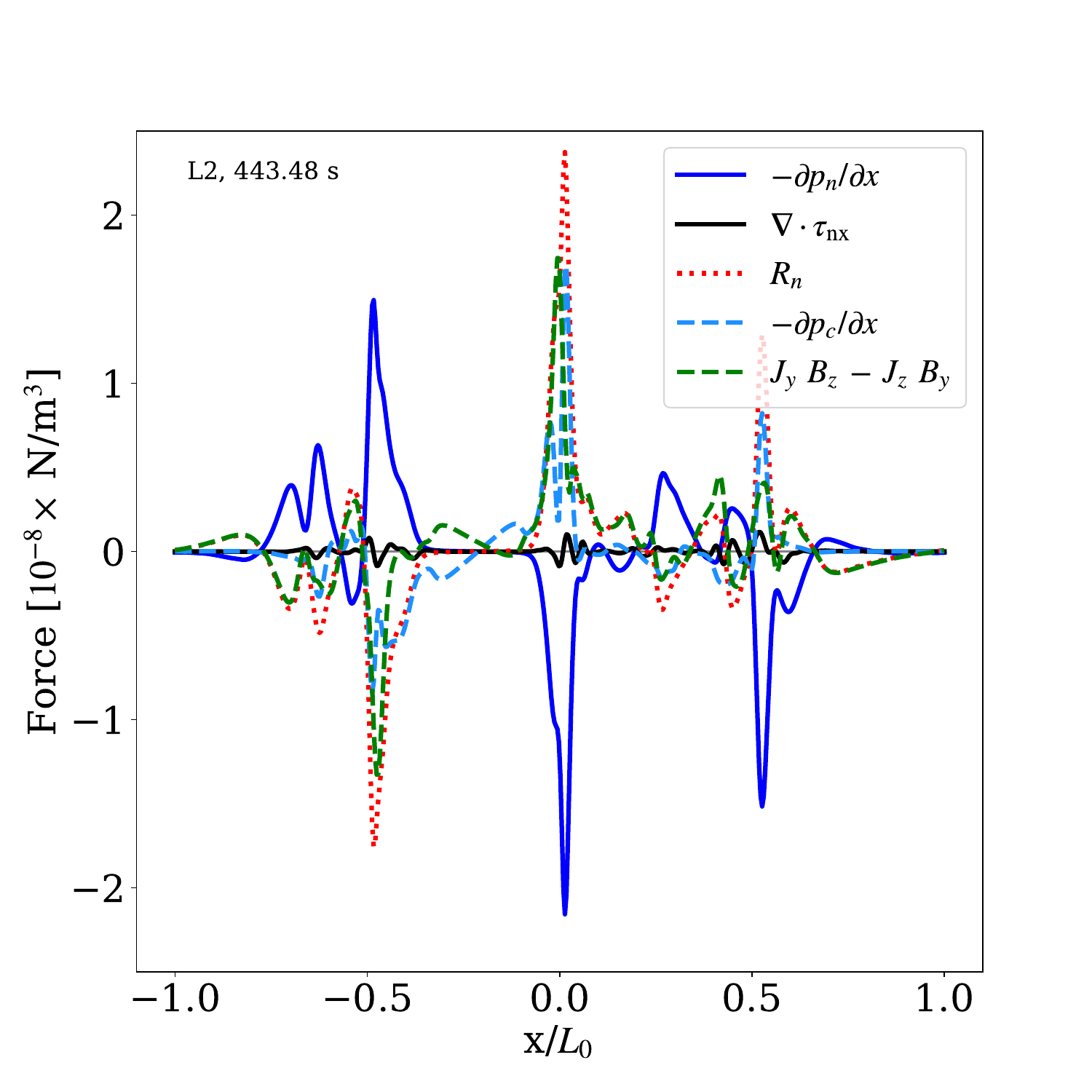}
 \caption{Forces that enter in the $x$-momentum equations at $z=0$ for neutrals only (solid lines):  pressure gradient $-\frac{\partial p_n}{\partial x}$ (dark blue) and  viscous force $\mathbf{\nabla}\cdot\tau_{\rm nx}$ (black), where $\tau_{\rm nx} = \xi_{\rm n}\left(2 \frac{\partial v_{\rm xn}}{\partial x},0, \frac{\partial v_{\rm xn}}{\partial z} + \frac{\partial v_{\rm zn}}{\partial x}\right)^{\rm T}$ with $\xi_{\rm n}$ defined by Eq.~A.9 in \cite{Popescu+etal2018}; for charges only (dashed lines): the pressure gradient $-\frac{\partial p_c}{\partial x}$ (light blue) and the Lorentz force $\hat{x}\cdot\vec{J}\times\vec{B}$ (green);  and the ion-neutral collisional coupling between neutrals and charges $R_{n} = \alpha \rho_{\rm n} \rho_{\rm c}\left(v_{\rm nx} - v_{\rm xc} \right)$ (dotted line), with the collisional parameter $\alpha$ defined by Eq.~A.8 from \cite{Popescu+etal2018} for the two simulations: P (left panel) and L2 (right panel).}
\label{fig:forces_snap}
\end{figure*}

\begin{figure*}[!htb]
 \centering
 \includegraphics[width=8cm]{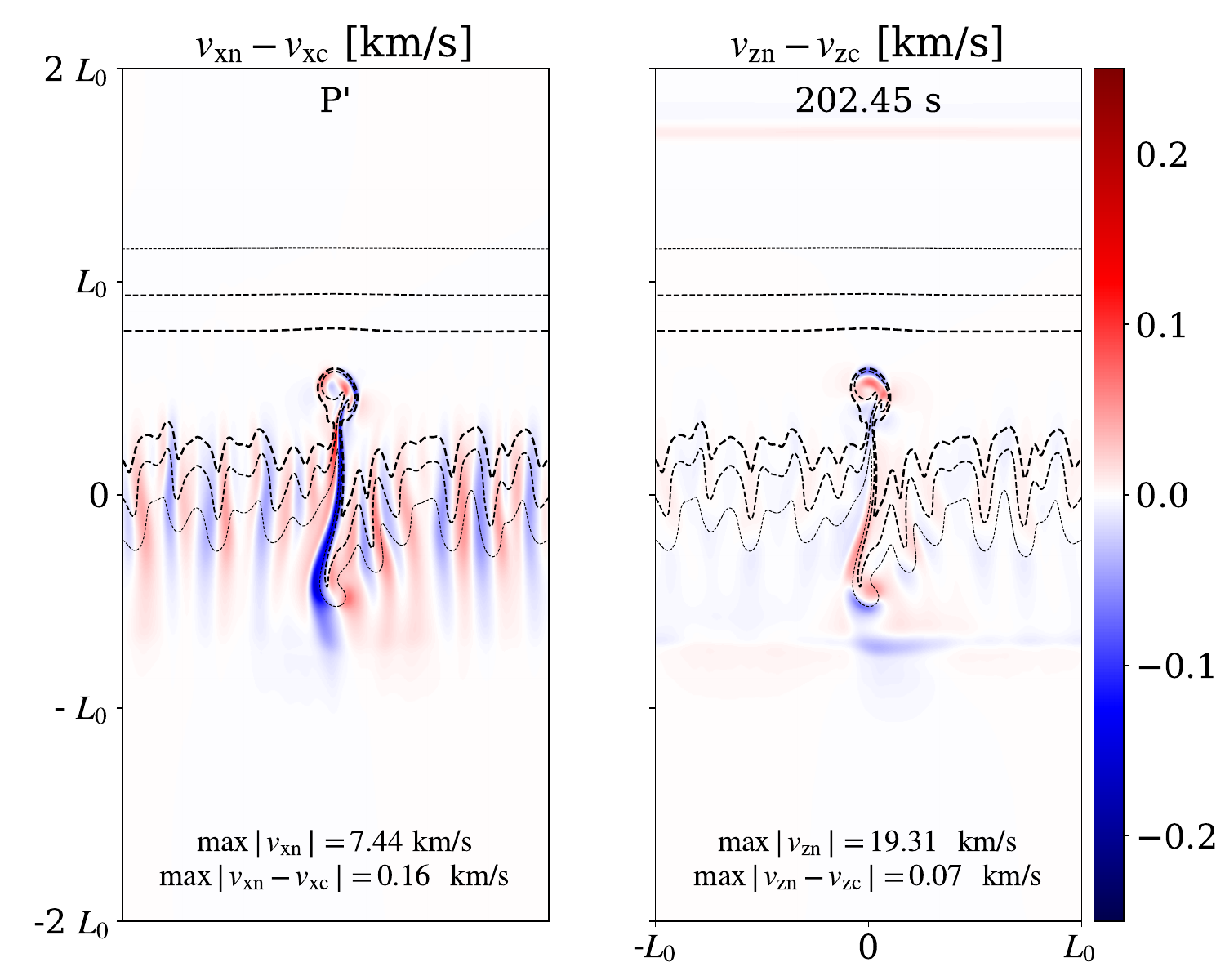}
 \includegraphics[width=8cm]{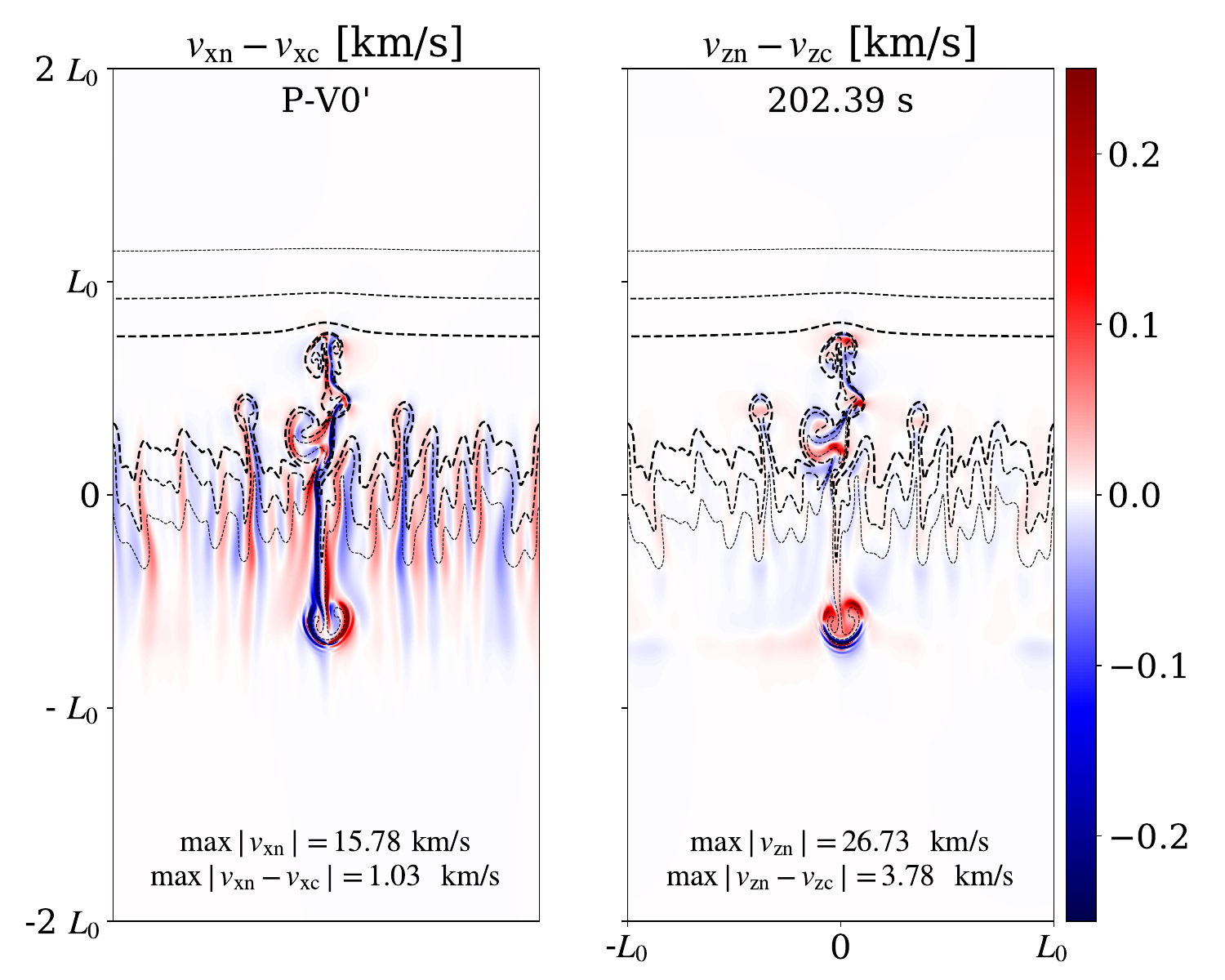}

 \includegraphics[width=8cm]{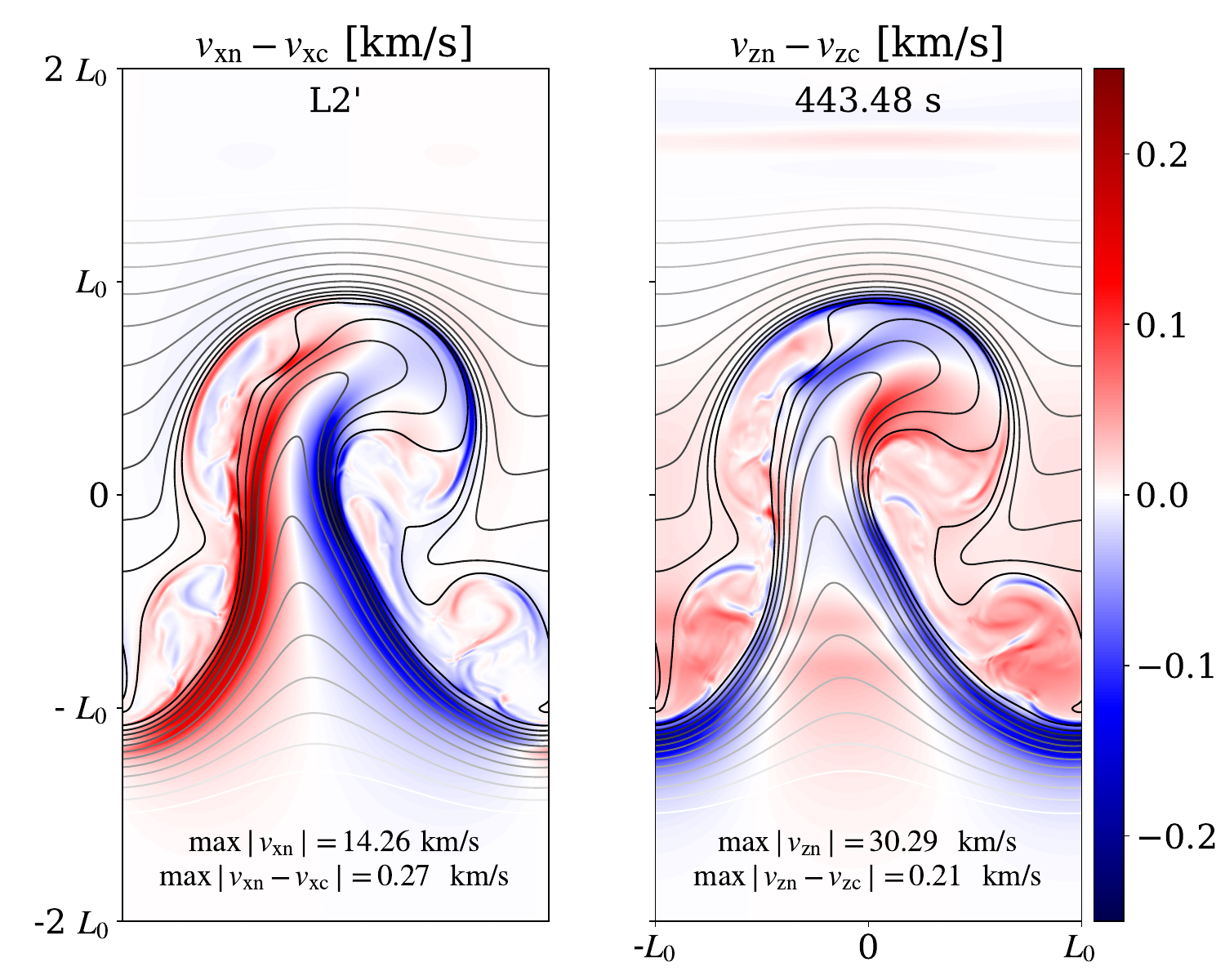}
 \includegraphics[width=8cm]{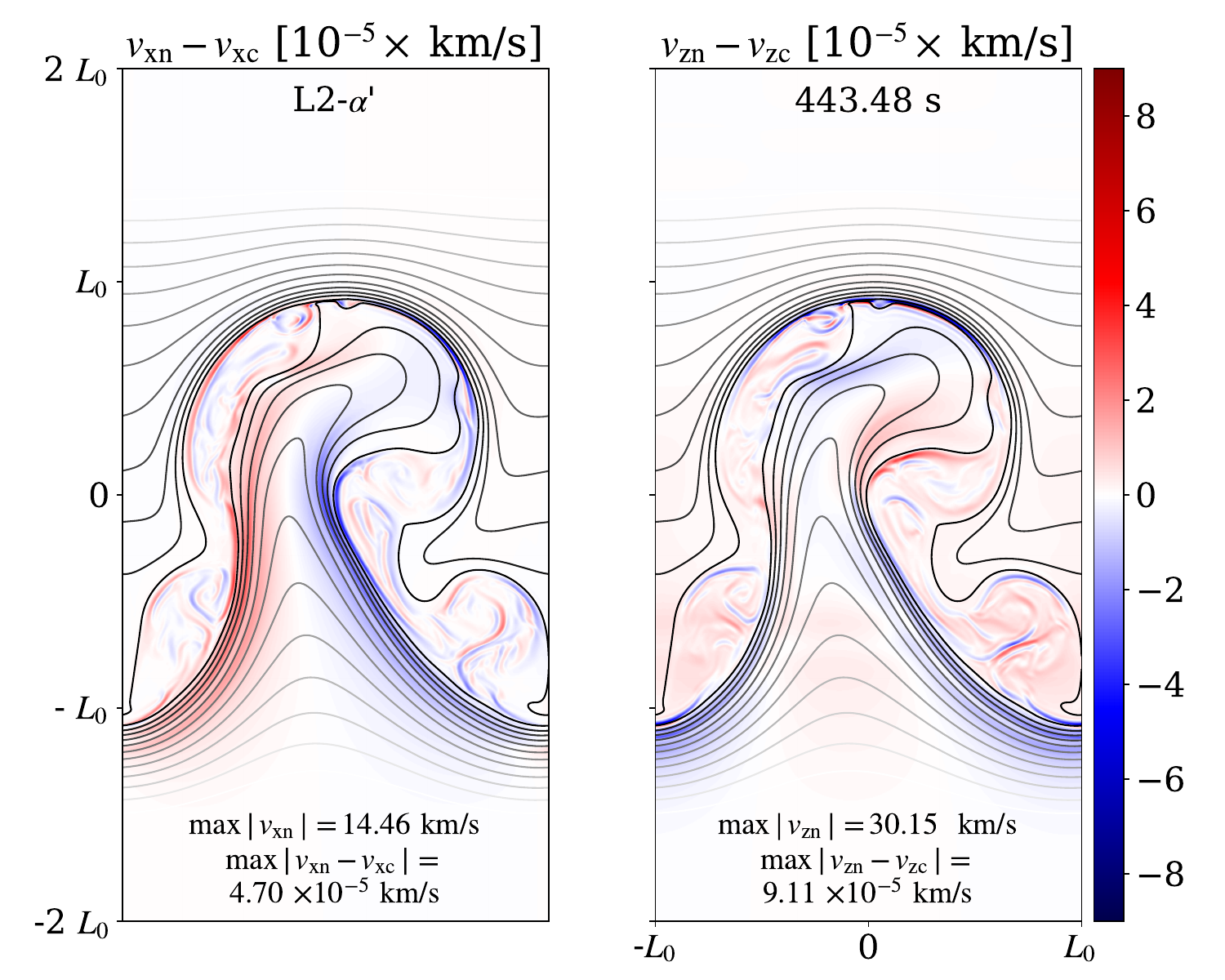}
\caption{{ Snapshots of the decoupling in horizontal and vertical velocities. Top left: ${\rm P}'$; top right: ${\rm P}$-${\rm V0}'$; bottom left: ${\rm L2}'$; bottom right: ${\rm L2}$-$\alpha'$, where the evolution of the background has been removed. The snapshots in the top row are shown at time $\approx$ 202 s, and those in the bottom row at time $\approx$ 443 s. The isocontours of the neutral density spanning the range from the minimum   to the maximum value are plotted for the snapshots of ${\rm P}'$ and ${\rm P}$-${\rm V0}'$ simulations (top). The magnetic field lines are plotted over the ${\rm L2}'$ and ${\rm L2}$-$\alpha'$ snapshots as isocontours of the magnetic potential (bottom), }with levels corresponding to values from 0.6$A_{\rm y}^{\rm max}$ and $A_{\rm y}^{\rm max}$, where $A_{\rm y}^{\rm max}$ is the maximum value of $|A_{\rm y}|$.}
\label{fig:dec_snap}
\end{figure*}
\begin{figure*}[!htb]
 \centering
 \includegraphics[width=16cm]{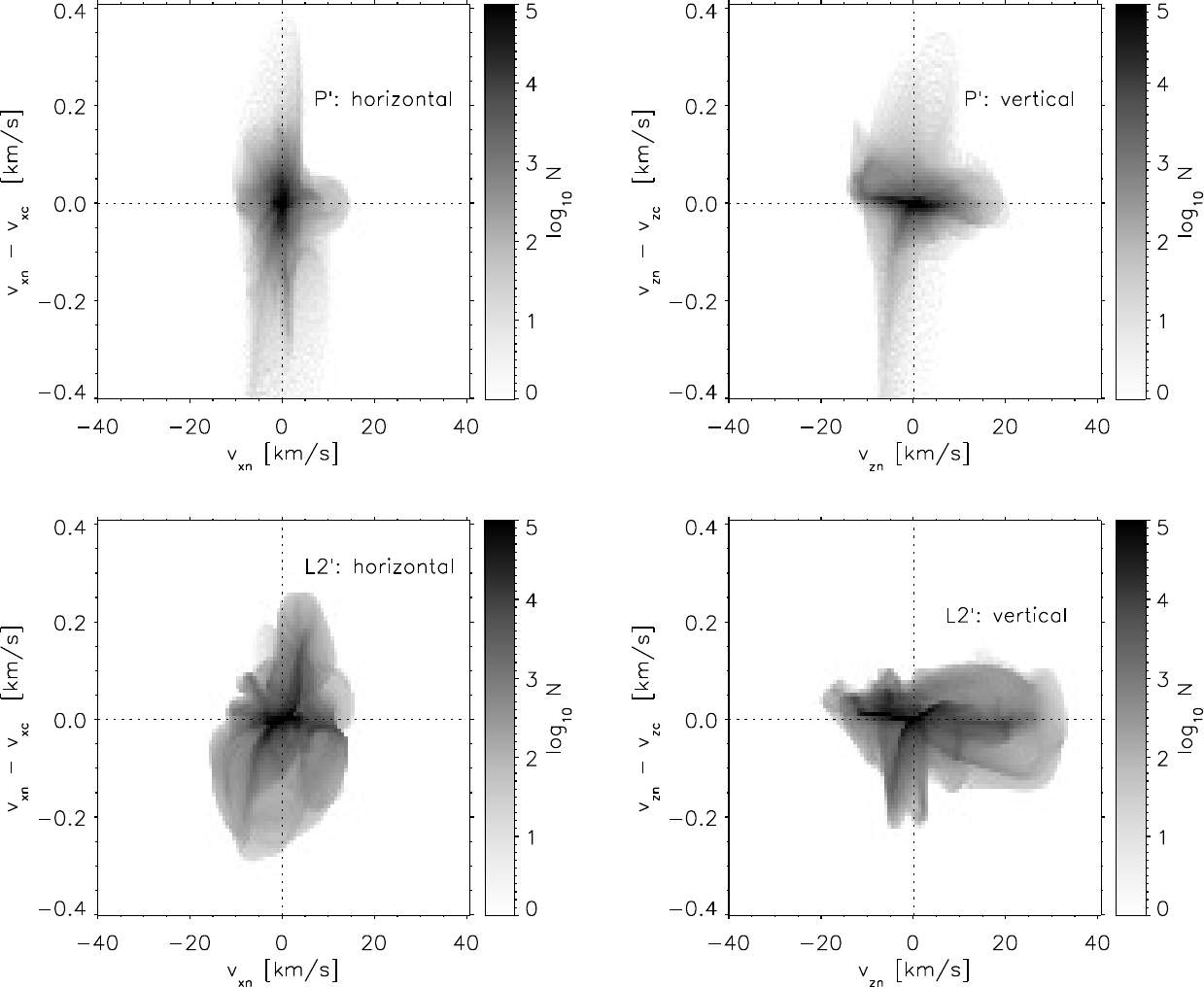}
\caption{
  Bivariate probability density functions between
  the  velocity ($x$-axis) and the decoupling in the velocity ($y$-axis)
   for the ${\rm P}'$ case (top panels) and ${\rm L2}'$ case (bottom panels), where the evolution of the background has been removed. 
   In both cases, all velocity values over the last 50 sec of each simulation, and at the vertical location between $-2L_0$ and $2L_0$ have been considered.  
    Left panels: Bivariate probability density functions corresponding to horizontal velocity;
   Right panels: Same, but for the vertical velocity.
The vertical dotted line  gives the location of zero velocity value.}
 \label{fig:dec_hist}
 \end{figure*}

We now quantify the magnitude of the ion-neutral decoupling in our simulations. We study the decoupling in velocities due to finite neutral-neutral and ion-neutral MFP effects during the nonlinear phase of the RTI for the perpendicular and sheared magnetic field configurations, respectively.  In order to focus specifically on the nonlinear RTI development, all figures in this section show simulated quantities with the corresponding evolution of the equilibrium subtracted.  This is denoted by the prime symbol: ${\rm P}' \equiv {\rm P} - {\rm P}_{\rm eq}$; ${\rm L2}' \equiv {\rm L2} - {\rm L2}_{\rm eq}$; ${\rm P\text{-}V0}' \equiv {\rm P\text{-}V0} - {\rm P\text{-}V0}_{\rm eq}$; ${\rm L2\text{-}}\alpha' \equiv {\rm L2\text{-}}\alpha - {\rm L2\text{-}}\alpha_{\rm eq}$.

To better understand how the decoupling is related to the flow, we first show the snapshots of the velocities for the  ${\rm P}'$ and ${\rm L2}'$ simulations in Figure~\ref{fig:vel_snap}. The flows for the simulations where the effect of the incomplete collisional coupling is removed, P-V0$'$ and L2-$\alpha'$, are very similar to the corresponding P$'$ and L2$'$ simulations, and the differences cannot be seen in the flow snapshots, hence they are not shown here. In order to highlight the location of the structures, the isocontours of the neutral density are shown for the P$'$ simulations and the isocontours of the out-of-plane magnetic vector potential are shown for the L2$'$ simulations.  
The peak velocity values appear to be related to the motion of the structures created by the RTI. This is especially evident in the L2$'$ simulation, where 
the flow boundaries are closely aligned with the magnetic field lines.

The decoupling in velocities between charges and neutrals appears when a force that acts on only  one of the two reaches the same order of magnitude as the coupling term between the two in the momentum equation. In our case the forces that act differently on the neutrals and charges are the electromagnetic forces that only act on charges, {the respective pressure gradient forces on each of the species, and the viscous force which is much greater for the neutral species. Thus, we expect the decoupling to be affected by the magnetic field evolution, strong relative density gradients, and the physical viscosity. 

To illustrate and quantify the different forces acting on the two species, Figure~\ref{fig:forces_snap} shows the contributions of the pressure gradient forces, the Lorentz force, the viscous force for the neutrals, and the friction force due to ion-neutral collisions in the $x$-direction at $z=0$ for the P$'$ and L2$'$ snapshots shown in Fig.~\ref{fig:vel_snap}.  The balance of forces is such that in both simulations the viscous force appears to make only a small contribution to the overall force balance, and the neutral pressure gradient is balanced almost exactly by the ion-neutral friction force in the momentum equation for the neutrals.  In the momentum equation for the charges the pressure gradient of the charges combined with the Lorentz force are similarly balanced by the ion-neutral friction force.  It should be noted that in the P$'$ simulation the Lorentz force and the pressure gradient of charges are always in the same direction, signifying that the purely out-of-plane magnetic field is simply advected along with the ionized fluid. On the other hand, the L2$'$ simulation shows regions where the Lorentz force counteracts the pressure gradient of the charges, leading to more complicated force balance dynamics for simulations with an in-plane component of the magnetic field.  It is also of note that the peak magnitude of the ion-neutral friction force, $R_n \approx 2.5\times 10^{-8}~\text{N/m}^3$, is several times the maximum initial gravitational force experienced by the prominence thread material, $g \rho_n^{\rm max} = (273.98~\text{m/s}^2) (1.834\times 10^{-11}~\text{kg/m}^3) = 5.025\times 10^{-9}~\text{N/m}^3$.}

Figure~\ref{fig:dec_snap} shows snapshots of the decoupling in the horizontal and vertical velocities for the simulations P$'$, P-V0$'$, L2$'$, and L2-$\alpha'$. This figure shows the neutral-ion velocity differences, $v_{\rm zn} - v_{\rm zc}$ and $v_{\rm xn} - v_{\rm xc}$, calculated at the same time $t\approx 202.4$~s for P$'$ and P-V0$'$, and $t \approx 443.5$ s for L2$'$ and L2-$\alpha'$. The same isocontours of neutral density and magnetic potential as in Figure~\ref{fig:vel_snap} are plotted over the snapshots of the decoupling to facilitate the comparison.
It is apparent that the decoupling is not small, reaching  a few hundred m/s, and in the P-V0$'$ simulation that neglects neutral viscosity up to several km/s. However, as it is highly localized, the average magnitude of decoupling over the domain is significantly smaller. Snapshots of the decoupling in the sheared field L2$'$ simulation reveal that the decoupling is strongest in locations of greatest magnetic field tension, where the magnetic field lines are bent and compressed. By comparing this figure to Figure~\ref{fig:vel_snap}, we also find a direct correlation between the local magnitudes of the flow and flow decoupling for both simulations, P$'$ and L2$'$. 

The impact of the finite neutral-neutral collision MFP on the degree of flow decoupling in the nonlinear regime is demonstrated by comparing the  P$'$ and P-V0$'$ simulation results.  We recall, as shown in Fig.~\ref{fig:time_snaps} and Fig.~\ref{fig_group_growing}, that the lack of viscous damping in the P-V0$'$ simulation results in stronger development of high-$n$ modes, and therefore more small-scale density and flow structure in the nonlinear phase of the RTI.  Comparing the top panels of Figure~\ref{fig:dec_snap} shows that this generation of small scales also leads to significantly enhanced decoupling between the ion and neutral flows.  We thus conclude that {despite a small contribution to the dynamical force balance,} proper accounting for the finite neutral-neutral collision MFP effects is necessary so as not to overestimate the expected flow decoupling as a result of nonlinear small-scale formation.


As  expected, the ion-neutral flow decoupling in the L2-$\alpha'$ simulation with a near-zero ion-neutral collision MFP (Fig.~\ref{fig:dec_snap}, bottom right) is five orders of magnitude smaller than the L2$'$ case (bottom left). This once again emphasizes the importance of accurately capturing the finite ion-neutral MFP effects.
{Another way to statistically quantify the magnitude of RTI flows and ion-neutral flow decoupling is via 
bivariate probability density functions between the velocities (x-axis) and the decoupling in the  velocities (y-axis)  within the simulation domain. The left panels of Figure~\ref{fig:dec_hist} show 
the bivariate probability density functions between the horizontal velocity  and the decoupling in the horizontal velocity for the P'  (top panel) and L2'  (bottom panel).
 The right panels of Fig.~\ref{fig:dec_hist} show
the same  for the vertical velocities.
These bidimensional histograms were  obtained over the last 50~sec of each simulation: for time intervals $t=\{393,443\}$s for L2$'$ and $t=\{200,250\}$s for P$'$, with the data taken over the central part of the simulation domain in the vertical direction, $z\in [-2L_0, 2L_0]$. 
One of the features of the bidimensional histogram of the vertical velocities in the L2$'$ simulation is a significant asymmetry with stronger flow in the positive direction.  This asymmetry can be attributed to the high velocity  values  within the central bubble in the L2$'$ simulation, as shown in Fig.~\ref{fig:vel_snap}. We also observe that vertical velocities are higher than horizontal velocities for both L2$'$ and P$'$, which is again consistent with the velocity snapshots in Fig.~\ref{fig:vel_snap}, and is due to the gravitational force driving the RTI flow. 

The bidimensional histograms show that the   decoupling is mostly symmetric in the P$'$ case. In the L2$'$ simulation there are indications of asymmetry in the vertical velocity differential, showing peak negative velocity differential to be twice as strong as the peak positive differential. 
This asymmetry is produced primarily by the neutrals falling down faster than the charges at the bottom of the spike (bottom left quadrant), while
the decoupling in the upflows is approximately symmetric.
This can also be seen in Figure~\ref{fig:dec_snap}, and is  due to the magnetic tension forces that act on the charges, but not the neutrals, in supporting the prominence material against gravity.
}


We observe that while the horizontal flow velocities are lower than the vertical values, the flow differential is more pronounced in the horizontal than the vertical direction.  This is particularly notable for the P$'$ simulation, most likely because the structures are long and thin, creating large horizontal density and pressure gradients leading to flow decoupling.  While the highly localized decoupling cannot be directly observed with the current telescope resolution, the histograms of the flow decoupling obtained from observations could be compared to the estimates based on computer simulations and analysis such as those presented in this paper.

\section{Conclusions and discussion} \label{sec:conc}

In this article we  explored the development of the Rayleigh-Taylor instability by focusing on the effects of elastic and inelastic collisions between and within neutral and charged particle populations in the context of a two-fluid model. 
We studied how the linear and nonlinear stages of the instability depend on the collisional MFP parameters, on their influence on the RTI mode growth rates, and on the decoupling between neutrals and charges. The conclusions of this work can be summarized as follows:
\begin{itemize}

\item The action of inelastic processes of ionization--recombination leads to the formation of high-density structures within the ionized fluid that forms at the interface between the cold prominence and the hot coronal material.  These structures follow the evolution of the RTI eddies, and are due to the ionization of the
neutrals at the interface. The RTI linear growth rate is not affected by   inelastic collisions.

\item The effects of finite elastic collision MFP for both ion-neutral and neutral-neutral collisions (e.g., viscosity) suppress the RTI linear growth rate on small scales.
Ion-neutral collisions qualitatively modify the structure of RTI modes for high mode numbers; the effect of viscosity appears at yet smaller scales.

\item During nonlinear RTI development, non-negligible decoupling between velocity fields of neutrals and charges, at a few hundred  m/s, is observed at locations of strong gradients in neutral density and/or magnetic field.  This localized decoupling is due to the differential between the MHD forces acting separately on charges and neutrals becoming comparable to the collisional ion-neutral coupling force. 
The decoupling is more pronounced on small spatial scales. 

\item Statistical analyses of the simulation results using histograms of the flow map demonstrate up-down asymmetry in both flow and flow decoupling, in particular in simulations with a sheared magnetic field configuration.  Furthermore, while vertical flow is generally stronger than horizontal flow, the ion-neutral flow decoupling is more pronounced in the horizontal direction for both magnetic field configurations.  These statistical features can potentially be detectable in observations.

\end{itemize}

As noted above, as a consequence of non-equilibrium ionization, we observe that the RTI instability leads to the formation of a relatively thin high-density layer of ionized fluid surrounding the falling cold high-density neutral prominence material.  This non-equilibrium ionization (and excitation) process is one possible explanation for the enhanced emission detected at the prominence corona transition region at the edges of the bubble in observations by \cite{Berger_2017}.
The excessive  brightness of the layer of high-ionization material seen at the boundary of the Crab Nebula \citep{prom3,prom2} could  similarly be explained by a higher electron density than otherwise expected.

A two-fluid model allows  the quantification of the degree of coupling between ions and neutrals. We found large values of the decoupling in velocity, which might be related to the magnetic activity being localized at the places where the magnetic field is significantly stressed. As the decoupling is highly localized, it is difficult to measure it directly from observations.  Direct measurements of the magnetic fields in coronal structures are similarly challenging.  However, statistical information about the decoupling, such as histograms, could be compared between the theoretical estimations and the observations to infer information about both plasma flow and local magnetic field structure.


\cite{Diaz2012} studied analytically the instability in a two-fluid approach where they consider a three-dimensional setup with two uniform regions separated by an interface, and a uniform magnetic field everywhere. They find a much lower growth rate in the two-fluid approach compared to the single-fluid approach, the reason for the decrease being that the ion-neutral interactions are not taken into account in the MHD approximation.
{ Our result is similar to that of  \cite{Diaz2012} in that the collisional interaction between neutrals and charges decreases the growth rate. }

\cite{2014bKh} performed a high-resolution numerical study of the RTI in a much more idealized setup. Similar to several studies of magnetic flux emergence through the solar photosphere (e.g.,  \citealt{Arber_2007} and \citealt{Siverio_2020}), they modeled partial ionization effects within a single-fluid approach via the ambipolar diffusion term. In agreement with the analytical study of \cite{DiazKh2013}, \cite{2014bKh} found an increase of about 50\%  in the growth rate on small scales in the simulations that included the ambipolar term  compared to pure MHD simulations. The increase in the RTI growth rate was observed on scales where in the single-fluid approximation the in-plane component of the magnetic field imposes a cutoff. When partial ionization effects are taken into account, it was shown
that the cutoff is removed, allowing for a small but non-negligible linear growth rate on small scales beyond the MHD cutoff. While this might appear to contradict the conclusions of the present work, there is a critical difference between the two in the choice of the RTI-unstable equilibrium.
This key difference is the more realistic smoothly varying interface between the cold dense prominence and the hot low-density coronal material considered here. 
Another source of the difference between the results of \cite{2014bKh} and the present work is that the MHD model with ambipolar diffusivity assumes more coupling between ions and neutrals than a fully two-fluid model and may capture their interaction in a different manner.

\begin{acknowledgements}
This work was supported by the Spanish Ministry of Science through the project PGC2018-095832-B-I00, by  KU Leuven through a PDM mandate and the US National Science Foudation. It contributes to the deliverable identified in FP7 European Research Council grant agreement ERC-2017-CoG771310-PI2FA for the project ``Partial Ionization: Two-fluid Approach''.
The
work was supported by the funding received from the European Research Council (ERC) under
the European Union’s Horizon 2020 research and innovation programme (grant
agreement No. 833251 PROMINENT ERC-ADG 2018).
{Any opinion, findings, and conclusions or recommendations expressed in this material are those of the authors and do not necessarily reflect the views of the National Science Foundation.}
The authors wish to acknowledge the contribution of Teide High-Performance Computing facilities and LaPalma Supercomputer to the results of this research. TeideHPC facilities are provided by the Instituto Tecnol\'ogico y de Energ\'ias Renovables (ITER, SA). URL: http://teidehpc.iter.es

\end{acknowledgements}

\bibliographystyle{aa}
%

\end{document}